# The Rise of Network Ecology: Maps of the topic diversity and scientific collaboration

Stuart R. Borrett [*,a,c], James Moody [b,c], Achim Edelmann [b,c]

[a] Department of Biology and Marine Biology
University of North Carolina Wilmington
Wilmington, NC 28403

[b] Department of Sociology
Duke University
Durham, NC 27708

[c] Duke Network Analysis Center
Social Science Research Institute
Duke University
Durham, NC 27708

* Corresponding Author, borretts@uncw.edu, 910-962-2411

## Abstract

Network ecologists investigate the structure, function, and evolution of ecological systems using network models and analyses. For example, network techniques have been used to study community interactions (i.e., food-webs, mutualisms), gene flow across landscapes, and the sociality of individuals in populations. The work presented here uses a bibliographic and network approach to (1) document the rise of Network Ecology, (2) identify the diversity of topics addressed in the field, and (3) map the structure of scientific collaboration among contributing scientists. Our aim is to provide a broad overview of this emergent field that highlights its diversity and to provide a foundation for future advances. To do this, we searched the ISI Web of Science database for ecology publications between 1900 and 2012 using the search terms for research areas of *Environmental Sciences & Ecology* and *Evolutionary Biology* and the topic tag *ecology*. From these records we identified the Network Ecology publications using the topic terms *network*, *graph theory*, and *web* while controlling for the usage of misleading phrases. The resulting corpus entailed 29,513 publications between 1936 and 2012. We find that Network Ecology spans across more than 1,500 sources with core ecological journals being among the top 20 most frequent outlets. We document the rapid rise in Network Ecology publications per year reaching a magnitude of over 5% of the ecological publications in 2012. Drawing topical information from the publication record content (titles, abstracts, keywords) and collaboration information from author listing, our analysis highlights the diversity and clustering of topics addressed within Network Ecology and reveals the highly collaborative approach of scientists publishing in this field. We conclude that Network Ecology is a large and rapidly growing area of ecology partly because the relational tools it rests upon are broadly useful for ecology, which is fundamentally a relational science. We expect continued growth and development of this research field.

"I map, therefore I am" Katharine Harmon (2004)

"Networks are everywhere" Manuel Lima (2011)

# 1 Introduction

Network concepts, tools, and techniques have a long history of use in ecology, but in recent years ecological science using networks appears to have grown rapidly (Ings et al., 2009, Proulx et al., 2005). For example, Lindeman (1942) first mapped the trophic interactions between species in Cedar Bog Lake thereby creating what today we would call a food web. In his 1968 recruitment lecture at the University of Georgia titled "The network variable in ecology", Bernard C. Patten explicitly recognized the importance of network models in ecology (Patten and Fath, 2000: page 178). More recently, scientists have used network models to investigate communities of mutualistic species (Bascompte et al., 2003; Bascompte and Jordano, 2007; Guimãres et al. 2011), general properties of ecosystems (Higashi and Burns, 1991, Jørgensen et al., 2007; Ulanowicz, 1986), and the movement of genes and organisms across landscapes (Holland and Hastings, 2008; Urban and Keitt, 2001). In this paper, we investigate the broad use of network concepts, tools, and techniques to investigate ecological and evolutionary questions. Following Borrett et al. (2013), we call this science domain *Network Ecology*.

Fundamentally, network models map a relationship(s) among a set of objects or actors (Higashi and Burns, 1991; Wasserman and Faust, 1994; Newman, 2003; Brandes and Erlebach, 2005). As such, they are a way of describing how objects are arranged with respect to each other that accounts for mutual dependencies and higher order characteristics in the resulting pattern. Analytically, networks can be represented as mathematical graphs ($G$) composed of a set of nodes or vertices ($V$) and edges ($E$) such that $G = (V, E)$. Both vertices and edges can have multiple characteristics such as different types or weights. Further, edges can be undirected (symmetric relation) or directed (asymmetric relations), which is visually indicated by arrows. For example, in a food web the vertices represent different species, functional groups of species, or abiotic resource pools like detritus, and the directed edges map the relationship "is eaten by" as represented by arrows pointing from prey to predator. Furthermore, vertices in a food web might be weighted by biomass or organism body size and edges might be weighted by the amount of energy or biomass transferred. In contrast, mutualistic networks are comprised of two kinds of vertices that represent two distinct classes of species (e.g., plants and pollinators). In this kind of network edges map mutualistic relationships as indicated by undirected lines that connect vertices of different, but never the same class.

Ecologists have used multiple types of network models. For example, ecologists interested in animal behavior and social structure have mapped the interactions among individuals of a population (Croft et al., 2004; Foster et al., 2012; Wey et al., 2008). At the community level, Ings et al. (2009) identify three broad types of ecological networks: food webs, mutualistic networks, and host-parasitoid networks. In the tradition of community ecology, this network classification scheme builds upon the nine possible qualitative interaction types between two species identified by Burkholder (1952) using a pairwise cross of positive (+), neutral (0), and negative (–) effects of one species on another such that mutualism between two species is indicated as (+,+). Ecosystem ecologists are interested in the same set of key relationships, but infer those relationships from transactive network models that trace the flow of a thermodynamically conserved tracer like energy or nutrients (nitrogen, phosphorus, etc.) through a given system (Fath et al. 2007). Such networks may represent food webs like in the community ecology networks (Cross et al., 2011; Martinez 1991), but they may also map non-trophic processes such as death and excretion (Baird and Ulanowicz, 1989; Olff et al., 2009; van Oevelen et al., 2011) or biogeochemical processes (Christain et al., 1996; DeAngelis, 1992; Reiners, 1986). Within the ecosystem approach, Scharler and Fath (2009) suggest that two schools of analyzing ecosystem networks have emerged. One school is based on the environment-centered work of Bernard C. Patten (Fath and Patten, 1999; Patten et al., 1976; Schramski et al., 2011) and the other school is based on the information-centric approach of Robert E. Ulanowicz (Ulanowicz, 1997, 2004). Although the approaches of both are based on Input–Output Analysis, they have led to largely independent yet braided theoretical developments. More recently, ecologists have been considering how to combine the variety of different ecological network perspectives to develop a broader understanding (Fontaine et al., 2011; Knight et al., 2005; Belgrano et al., 2005). From these examples, we might infer that network models and analytical tools have been used broadly in ecology. The question is how broadly?

Our objective in this study was to identify, map, and characterize the domain of Network Ecology. We addressed three primary questions. First, how large is the domain? Second, what topics are ecologists addressing using the network approach, and third what is the nature of the scientific collaboration among these ecologists? The examples previously presented suggest that there are a large number of disconnected topics being studied by a fragmented community of scholars. This would lead to a broad set of topics and a fragmented or highly clustered community. As publications are a key product of the scientific process, we used a computational approach to infer from the publication record both the primary topics in the field and the structure of scientific collaboration. This bibliographic approach draws topical information from the content of publication records (titles, abstracts, keywords), collaboration information from author listing, and prominence and subfield information from citations. While our approach is different from a traditional in-depth review of the literature, it serves as a broad and high-level review of the field that provides a foundation for future work.

# 2 Materials & Methods

To address our research questions, we used a combination of bibliographic techniques and network modeling. Similar network approaches to bibliometric studies have been used successfully to characterize the social structure of collaboration in many fields, including sociology (Moody, 2004), physics, biomedical research, and computer science (Newman, 2001a, 2001b), and the study of ecosystem services (Costanza & Kubiszewski, 2012). Topical modeling for science-studies is similarly widespread, mapping detailed portraits of particular fields (Moody and Light; 2006; Evans and Foster, 2011; Borner, 2010).

## 2.1 Bibliographic Data: Search and Selection Criteria

To identify the Network Ecology publications, we searched the ISI Web of Science (WoS). We chose this bibliographic database because it is the largest general index for science that also includes extensive indexing of ecological science. Within the WoS, we limited our search to the Science Citation Index Expanded and Social Science Index Expanded citation databases between 1900 and 2012. Given our focus on the science of ecology, we excluded the Arts & Humanities Citation Index. Furthermore, we excluded two conference proceedings databases because they cover a smaller period of time (1990-present) and because the discipline of ecology values journal article publications more highly than conference proceedings.

Network ecology lies at the intersection of (A) ecological science and (B) network concepts, tools, and techniques. To identify the broad domain of ecological science in the WoS, we searched the union of two ISI research areas, *Environmental Science & Ecology* and *Evolutionary Biology*, and the topic tag of *ecology*. This is broadly inclusive of ecological science, but it is dependent on the ISI research area classifications. To find the network science within this ecological science, we further searched the results for the union of three topic terms: *network*, *graph theory*, and *web*. These terms are frequently used within network science and can be regarded as signals for research using network concepts or models. The use of *web* captured science from spider webs to food webs, but it also included some spurious references to topics like the World Wide Web.

Ecologists use the term *network* in a variety of ways. The introductory examples illustrated the construction and analysis of models to characterize communities and ecosystems. However, there are common uses of the term network that are less related to the type of network science we sought. To address this challenge, we excluded those records that our initial search identified only due to their use of a few selected phrases. Specifically, we excluded records that were initially included only because they entailed the phrases *neural network* or *Bayesian belief network*. These phrases are common statistical techniques unrelated to the broader network science that was our target. We also excluded records identified solely on the term *monitoring network*, as this is often used to refer to a collection of sampling sites with ill-defined relations among the sites. We further excluded records that appeared due to *transportation network* and *railway network* because these phrases did not typically recover research where network science was applied to ecological problems.

We downloaded the final selection of records from WoS on May 13, 2013. We analyzed the resulting corpus with network analysis tools in SAS, Pajek, and R.

## 2.2 Analysis

### 2.2.1 Publication Volume

Given the Network Ecology corpus, we characterized the volume of the publications using several metrics. First, we examined how the number of publications changed through time. To evaluate if the observed trend is driven by the more general publication inflation in ecology, we normalized the raw number of Network Ecology publications by the number of publications each year in ecology (WoS search for ecology as described previously without applying the search for network concepts). Our second approach for characterizing the publication volume was to quantify three aspects of the publications in the corpus. We described the distribution of (1) the ISI research area categories, (2) the journals in which the papers are published, and (3) the article citation frequency within the corpus. We describe both the richness and evenness of these distributions and list the most abundant contributions.

To evaluate the relative importance of the Network Ecology publication volume in 2012, we compared it to the relative frequency of selected concepts and tools in the broader ecology corpus in WoS. We started with the 50 concepts that were identified by members of the British Ecological Society (BES) as important ecological concepts in 1986 (Cherrett, 1989), generalizing when appropriate, and added 13 additional terms to introduce selected, new terms of apparent importance (e.g. *disease*), and common analytical techniques not included in the BES list (Table 1).

Table 1 Goes Here

### 2.2.2 Corpus Validation

To check the quality of the Network Ecology corpus discovered we determined if the corpus included recent publications considered important by a community of experts. Participants were surveyed online and asked to identify up to 5 Network Ecology publications between 2007 and 2012 that they considered most important (see Table S1). The survey was created using the Qualtrics software and distributed to the Ecological Society of America's Ecolog-L electronic mailing list (hosted by the University of Maryland), which currently boasts more than 17,000 subscribers. We also sent the survey directly to a targeted group of 56 known Network Ecology experts. A reminder was sent out after about a week and participants were encouraged to further distribute the survey link. The resulting convenience sample was meant to provide a positive control on our discovery of key Network Ecology publications. Accordingly, we classified the references identified by the expert community into three categories: (1) not included in WoS, (2) included in WoS but missing from the corpus, and finally (3) included in the corpus.

### 2.2.3 Topics in Network Ecology

To identify the predominant topics in Network Ecology, we built a network of papers linked by similar terms used in the title, abstract, and keyword records. Edges are weighted by the similarity of their term use, using the standard tf-idf formulation (Börner et al., 2003), which discounts the similarity of common terms and favors more rare terms. Edges were included in the topic network when they indicated a minimum percentage of co-term similarity. We constructed two versions of this co-term network; one composed of all papers in the corpus (35% minimum similarity) and a second with 5-year moving windows (25% minimum similarity). By this construction, papers on similar topics will "hang together" in the network which can then be identified using cluster detection techniques (Moody and Light, 2006). For this purpose, we used the Louvain community detection algorithm (Blondel et al., 2008) on a weighted graph as implemented in the PAJEK software. We constructed two-dimensional maps of the topic space by applying space-based layout routines to the similarity network. These routines place papers that have much in common near each other. Since the network representation of such maps are too dense to be substantively useful, we constructed contour maps that reflect the density of papers in each region of the topic space and label these maps with the most frequent terms used in each cluster (Moody and Light, 2006 for this particular technique; Börner 2010, for general approach).

In addition to identifying the clusters within these waves, we characterized the topic distribution using network community detection metrics. The first is the *modularity score* (Newman and Girvan, 2004), which captures the extent of clustering beyond random chance. A value of 1.0 would indicate completely disconnected clusters, while a value of 0 would be no different from random assignment. The second is the *heterogeneity index*, which captures the distribution of topics across clusters. The heterogeneity index is the probability that two papers chosen at random would fall within the same cluster. Labels are generated automatically using the most commonly used terms within each cluster.

#### 2.2.4 Structure of Scientific Collaboration

To characterize the structure of scientific collaboration in Network Ecology, we focused on article co-authorship. We constructed a co-authorship network in which nodes represent individual authors, and weighted edges connect two authors by the number of papers they have co-authored. Since publication records are often inconsistent in name use we applied a light name-cleaning routine to combine names that are obviously similar ("Stuart J. Whipple" & "S.J. Whipple"), by matching on uncommon last names and combinations of first and middle initials (see Moody, 2004). This is a deliberately conservative name correction routine, as the network costs of conflating common names (potentially creating a bridge between groups that are unconnected) are typically worse than leaving them separate (which while potentially increases isolates tends to preserve collaboration groups).

Once constructed, we identify the connectivity structure and examine diversity in collaboration groups. We again applied the Louvain community detection algorithm to identify collaboration clusters. Here, we used the default unit-weighted resolution parameter.

# 3 Results

## 3.1 Publication Volume

The total number of Network Ecology articles discovered prior to any exclusions based on selected phrases was 33,900 (Table 2). The majority (59%) of these records were discovered by the intersection of the ISI research area *Environmental Science & Ecology* and topic term *network,* and 88% of the articles were discovered by the intersection of topic terms *network*, *graph theory*, and *web* with the ISI research area *Environmental Science & Ecology*. In contrast, the intersection of these three network topic terms with the ISI research area *Evolutionary Biology* only found 8% of the records and only 4% of these were unique. Adding publications identified by the topic *ecology* to define ecological science added 2,832 unique records to the corpus (8%). Notice that percentages do not necessarily add to 100% because a record may have multiple classifications and may include more than one of the key words. From this initial corpus, we excluded records based on their use of phrases that entail the term *network* but do not necessarily refer to research that applies network science to ecological problems (# affected): *neural (2,444)*, *Bayesian belief (259)*, *monitor (1,263)*, *railway (232)*, and *transport network* (244). Combined, these exclusion phrases remove 3,962 records or 12% of our initial sample. The remaining 29,513 records are the Network Ecology corpus that we analyzed further.

Table 2 Goes Here

The number of Network Ecology articles published each year has grown rapidly (Fig. 1). The first ecology article discovered by our methods appeared in 1936, by Griswold and Crowell on "The Effects of Humidity on the Development of the Webbing Clothes Moth"; however, this article seems to have been identified due to the common name of the organism rather than its use of a formal network approach. Instead, Gimingham's (1961) article looking at the network of variation in heath communities appears to be the first ecological publication identified by our search that explicitly uses a network model. Since these early publications the number of Network Ecology publications increased to 3,063 articles in 2012. This growth in Network Ecology research cannot be explained by the general increase in number of ecology publications.

Relative to the total number of ecology publications, Network Ecology publications noticeably increased in the early 1960's (Fig. 1b). Between 1990 and 1991 the percent of Network Ecology articles jumped by almost 1%. Following this jump, Network Ecology publications have steadily increased from less than 1.5% in 1991 to over 5% in 2012. These developments show that Network Ecology is a large and rapidly growing area of research.

Figure 1 here

The Network Ecology corpus includes work from a wide range of ISI research areas and publication sources, which shows the diversity of research in this field (Fig. 2). Our corpus entailed publications pertaining to 122 unique, though non-exclusive ISI research areas. The most common area was *Environmental Science & Ecology*, which is not surprising as this was one of our search criteria. The next three most frequent research areas were *Marine & Freshwater Biology*, *Engineering*, and *Water Resources*. Other ISI research areas represented in our corpus included disparate disciplines such as *Geography*, *Urban Studies*, and *Public Administration*. These results indicate the broad character of research conducted in Network Ecology.

Figure 2 here

Network Ecology publications in our corpus were distributed across nearly 1500 different publication outlets. These are primarily journals, though sometimes proceedings from conferences.[1] The journals Marine Ecology Progress Series and Water Resources Research contained the most Network Ecology papers, but each only contained between 3 and 4% of the articles (Fig. 2b). Traditional ecology journals such as Ecology, Ecological Modelling, Oikos, Oecologia, Ecological Applications, and Ecology Letters were among the top 20 most frequent sources. However, so were more general journals such as Environment and Planning A, Science of the Total Environment, and Proceedings of the Royal Society B: Biological Sciences.

The frequency of citations to articles in the corpus (Fig. 3) roughly follows the expected long-tail distribution of a Zipfs or power-law form; however, there were fewer citations at the low-end than is typical for such distributions (Zipf, 1949). The modal value is 0 (13.8%) while 51% of papers are cited 8 or more times, 25% more than 23 times, 10% more than 50 times, and the top 5% received more than 78 citations. Only 3.3% of papers are cited 100 times or more.

Figure 3 here

In 2012, the Network Ecology publications listed in WoS were 5.1% of the total publications in ecological science as defined above (Fig. 1b). To put this proportion into perspective, we found the percentage of ecological publications in 2012 that contained other key ecological concepts including the 50 identified in the 1986 BES survey (Cherrett, 1989). The respective proportions range from 26.7% for *species* and 26.3% for *model* to approximately 0% for terms like *island biogeographic theory*, *pyramid of numbers*, and *3/2 thinning law* (Table 1). Some of the 1986 BES concepts continue to show relevance in 2012. For example, 17.5% of the ecology publications in 2012 included the term *population*, which represents a generalized form of the concept *population cycle,* which ranked fourth in the BES survey. *Ecosystem* was the highest ranked concept on the BES survey and it appeared in 10.7% of the ecology articles published in 2012. When compared to the relative frequency of these 63 terms, our definition of *Network Ecology* (which includes food webs) would have a rank of 15.

---

[1] Note that we do not correct here for merging or changes in journal names over time.

## 3.2 Corpus Validation

The online survey was completed by 59 people who identified 118 unique publications. Of these, 22 fell outside our target time frame (2007-2012). Of the remaining 96, 5 (5%) were not included in the WoS database. A closer look reveals that 3 were books or book chapters (Olesen et al., 2012; Whitehead, 2008; Ulanowicz, 2009), which are not indexed by WoS, and the other 2 were published in journals not indexed by WoS (Rudnick et al., 2012; Ulanowicz, 2011). For the remaining 91 publications, corresponding references could be found in the WoS database, and thus could, in principal, have been selected in constructing a reference corpus. After applying the search and selection criteria described above, 55 (60%) references identified by the experts were represented within our Network Ecology corpus.

Respondents showed agreement on some of the most important papers as multiple people identified the same papers. Out of the 96 publications identified, 13 (14%) were mentioned twice and 2 (2%) references were mentioned three times (Table 3). Despite the small scale, our survey suggests that our corpus captures the majority – but not all of the recent publications our convenience sample of Network Ecology experts considered important.

Table 3 here

## 3.3 Topics in Network Ecology

### 3.3.1 Corpus Topic Network

To identify the topic structure of the Network Ecology corpus, we first constructed a similarity network based on co-word frequency. Figure 4 illustrates this process by walking out from a single paper to its local neighborhood and the clusters they link to. In panel (a), we plot just the focal paper and its nearest neighbors. At an edge-threshold of at least 35% co-term similarity, this paper is directly adjacent to 15 other papers, all generally on aspects of food webs and grazers. Stepping out two more links as shown in panel (b), we reach nearly 300 papers and find this small cluster embedded within a wider field of 7 or 8 clusters, ranging in topics from algae in alpine environments to the invasive consequences of crayfish. Since the layout algorithm pulls similar papers near each other, these clusters emerge as tight “knots” in the network diagram. Many papers here are related in one way or another to water-based ecosystem studies, and were we to step out even further we would likely find those nested within a broader set of papers related to aquatic environments. To ease recognition, we visualized clusters by 2-d contour overlays as indicated in the example.

Figure 4 here

The complete topic network is comprised of 29,513 vertices representing the papers in the corpus connected by 106,795 edges indicating topic co-word similarity with a minimum of 35% similarity. This topic network is comprised of several components of varying sizes. The largest component contains 21,636 vertices (73% of total). The remaining 27% of the papers appear in smaller components. The second largest component contains 15 papers (0.05%). The majority of papers not in the largest component (5,908 or 20% of total) are isolated nodes (components of size 1) with no edges connecting them to other papers. This isolation indicates that either their topics do not appear related to other papers in the corpus or that their WoS records did not contain enough information to identify the similarity. We focused our subsequent analysis on the giant component of this topic network.

Figure 5 provides a contour diagram of the topic structure revealed in the giant component of the topic network. As in the previous example, vertex distribution in this space is based on a 2-dimensional layout of the underlying similarity-weighted network, so clusters of papers on similar topics will be placed near each other. Since a point-and-line graph of such a large network is visually indecipherable, we fit a 2-d kernel density surface to the distribution of papers in the space, creating a contour map of the network (Fig. 5). Topics are identified using the Louvain clustering routine, which optimizes assignments of vertices to clusters based on the relative weight of within-cluster compared to cross-cluster edges. The method automatically determines the number of clusters. We labeled the regions by the most common terms of the largest clusters found in that portion of the graph, with font size corresponding to frequency.

Figure 5 here

The overall topology of this topic network resembles a ring structure with multiple topic-centered "peaks" joined at their peripheries to a neighboring subfield. Starting at the north central region of the contour diagram, we find a large number of clusters generally related to aquatic ecosystems, rivers, and lakes. Moving in a clockwise direction, we then encounter a small ridge of work related to soil, nematodes, food web and communities that then links to work on predator–prey food webs and communities. The southeast of this map is composed of work on landscapes, habitats and conservation with a large genetic cluster. The common theme of land-use bridges from the southeast corner to the southwest which shows work on the intersection of human activity and ecosystems, particularly urban planning, policy & energy use. The far west of the diagram relinks with water system models through groundwater and distribution topics. Network statistics of the five largest clusters are indicated in Table 4.

### 3.3.2 Temporal Waves

To identify possible temporal variation of topics and topic structure in Network Ecology, we constructed and analyzed topic networks in temporal waves. We combined the years 1980 to 1989 due to the small number of papers. From thereon we constructed waves in 5 year moving windows. Table 4 shows the 5 largest topic clusters in each of the waves. Except in the years 1980 to 1989, the modularity score as well as the heterogeneity index shows remarkable stability across the waves. The modularity score as well as the heterogeneity index for the full corpus is substantially larger than that for each individual wave, which signals that the topics addressed did change across the wave as either new topics emerged, old topics disappeared, or the terminology used to address the same topic changed (our analysis cannot differentiate between these possibilities). Not only the size of the largest component, but also the average cluster size steadily increased from 38 for the years 1980 to 1989, to an average of 238 between 2005 and 2009 – a level that was subsequently almost researched within only 3 years (2010–2012).

Table 4 here

Our analysis reveals a strongly clustered pattern of the topic networks. The modularity score is over 0.82 in all waves since 1990, and 0.927 for the full corpus. This strong clustering is expected in a topical network for science studies, since most publications are designed to speak carefully to a particular research problem. The distribution of cluster sizes suggests that no single topic dominates the field of Network Ecology; the largest clusters typically account for between 5% and 7% of the topics in a wave. The topic network for the full corpus is even more dispersed with the largest clusters entailing only about 4% of the papers. The network for the full corpus allows us to identify more specialized topics that might not appear in any of the smaller waves as there would be too few papers. The more dispersed nature of the full corpus is reflected in the high heterogeneity scores, typically topping out over 0.95 in later years. Since all of these papers have a foundational commonality in Network Ecology, this suggests an extremely diverse body of topical work employing network concepts, tools, and techniques.

While the sheer diversity of the topical networks makes identifying temporal trends difficult, we did find some patterns of interest. We see similarities across the largest topics in each year as themes related to water and aquatic systems, predator-prey models, conservation, and network methodology reoccurs. In particular, topics related to predator-prey models reappear throughout most waves. Furthermore, topics related to water and aquatic systems (*lake*, *fish*, *phytoplankton*) seem to have become more central between 1995 and 1999 and gained further relevance thereafter (*stream*, *river*, *water, rain*). Between 2000 and 2004 conservation and the influence of human living and production on the environment become apparent (*reserve, pollution*, *air*, *concentration*, *road*, *emission,* and eventually also *firm, innovation*) which might be due to a more general interest in sustainable living and environmental protection. Between 2010 and 2012, themes related to urban living (*urban, city*) and the topic *temperature* appears, which could be due to an increased interest in climate change.

## 3.4 Scientific Collaboration

### 3.4.1 Overview

To map the structure of scientific collaboration in Network Ecology, we constructed a co-authorship network from the publications in our corpus. Whenever two scientists collaborate on a paper, it creates a connection that extends through all co-authors and thus collaboration networks provide a useful model for communities of science. In Network Ecology collaboration is common (Fig. 6). Most of our papers contain multiple authors and the median paper has 3 authors (inter-quartile range (IQR) 2 to 4), though there is a fairly large tail (with the largest paper having 82 authors). The incidence of collaboration has been growing over time, from an average of 2 in the 1980s to over 4 currently.

Figure 6 here

Concatenating across all publications we can build the full collaboration network. In this network, the vertices represent authors that are connected by an edge if they have co-authored a paper in the corpus (co-authorships among these authors on papers other than those in our corpus are not considered). Edges are weighted to indicate the number of papers co-authored. The corpus contains 69,564 uniquely named authors. Figure 7 provides an image of the full collaboration network using the same contour overlay strategy as previously described. The largest connected component contains 46% of these authors and only 2,695 (~3.9%) of the authors appear as isolated nodes having not co-authored a publication. The component size distribution is skewed, such that the next largest component has 75 people. The diameter of the large component is 27 steps, while the average path length is 7.66. Within the largest component, the largest bi-connected component contains 58% of authors (19,015), and 85% of this set are members of at least a 3-core – sharing at least three neighbors in common with every other member of the set. It is possible for most of the members of the largest connected component to reach each other via multiple collaboration paths. Thus, at a high-level of analysis, these results show a well-connected research community

Figure 7 here

### 3.4.2 Clustering

Despite the broad connectivity within the community, there are a large number of co-authorship clusters (Fig. 8). The Louvain community detection algorithm (Blondel et al., 2008) identified 149 clusters ranging in size from 6 to 1,618. The small clusters (<25 or so) are generally "fringe" cases that are only weakly connected to the rest; looking at only those with larger sized communities the median is 190 people (IQR: 87-354). The clusters congregate around three distinct larger communities, evident as peaks in the overall sociogram (Fig. 7). Table 5 provides the three authors with the highest betweenness centrality within the largest clusters. At this scale of analysis, these clusters represent groups of authors who generally work together on similar scientific problems or topics. Embedded within the clusters are also traces of collaborative working groups that may be based on a single primary investigator or a small group of collaborative principles. For example, G. Woodward, J. Memmott and N. Martinez have the highest betweenness centrality in cluster 13 (Table 5), which contains 1,618 authors (about 2.5% of all authors, 5% of the largest component). These authors clearly work on a variety of topics, but they appear to also have a common focus on traditional community networks like food webs and pollination networks (Woodward et al. 2005, Woodward et al. 2008, Martinez 1991, Williams and Martinez 2000, Memmott 2009). Furthermore, several authors within this cluster have collaborated on recent synthetic publications (Brose, 2006; Ings et al., 2009).

The modularity score for the overall clustering of the co-authorship network is 0.92, suggesting relatively distinct clusters. Just over 94% of co-authorship occurs within clusters, and 79.89% of authors have only within-cluster ties. The remaining 20.11% of authors create the links that pull together the entire network. Within cluster density averages 0.09 (IQR: 0.02 – 0.12), but since density is strongly affected by cluster size average degree is perhaps more telling. On average, authors collaborate with 9.2 other authors in their clusters (Median: 7.8l; IQR: 6.3 – 9.6). The structure within clusters is composed of overlapping cliques formed by sharing authorship on papers. This ranges from very fragile structures where a few key nodes chain across multiple large papers to very robust groups that collaborate across many papers.

Table 5 here

Figure 8 here

#### 3.4.3 Ego Networks

We can further highlight features of the Network Ecology collaborative structure by focusing on select individual authors. Figure 9 shows the ego co-authorship networks for four selected scientists: Bernard C. Patten, Robert E. Ulanowicz, Stephen R. Carpenter, and Derek C. G. Muir. These ego graphs show the co-author structure from the perspective of a specific individual. Here, we have also included the co-authorships that most directly link Patten and Ulanowicz to visualize the hypothesized historical separation between their research programs. Within the Network Ecology corpus, Patten has 62 direct co-authors, Ulanowicz has 67, and Carpenter and Muir have 128 and 226, respectively. These authors clearly have differently sized collaborative structures.

Each author's ego net tells different stories. Unpacking the structure of those ego nets highlights the different factors that can influence the structure of scientific collaborations. For example, Patten's ego net displays a couple of different working groups. The first is comprised of his former Ph. D. students (e.g., S. J. Whipple, S. R. Borrett, S. R. Schramski) and the Systems Ecology and Engineering colleagues at the University of Georgia (e.g., D. K. Gattie and C. Kazanci). Another work group is comprised of S. E. Jørgensen, and M. Straskraba. Brian D. Fath appears to be loosely part of both of these groups as well; he was also a Ph. D. student with Patten. Another cluster of co-authorships appears in the lower region of the plot and includes P. G. Verity, M. E. Frisher, and J. C. Nejstgaard. The publications that link these co-authors were a result of an NSF Biocomplexity award (Nejstgaard et al., 2006; Whipple et al., 2007). There is another cloud of co-authors plotted between Patten and Ulanowicz that is a result of a synthetic publication calling for improvement in food web construction co-authored by multiple investigators working on food webs at that time (Cohen et al., 1993).

Figure 9 here

# 4 Discussion

Considered together our results highlight three key features of Network Ecology. First, Network Ecology is a large and rapidly growing area of ecology. Related papers are published in a wide variety of journals, touch on a breadth of topics, and co-occur in a broad set of research areas defined in WoS. Second, while there is some continuity over time visible, the substantial research topics addressed within Network Ecology are manifold and have varied over time as shown by prominent co-occurrences of words across papers. Third, collaboration as reflected in co-authorship among researchers suggests a highly collaborative science, but one that appears fragmented with noticeable clusters of collaborators. We next discuss these results in light of (1) the forces that have drive the increase in Network Ecology, (2) the connections between network, systems, and ecological science, (3) the limitations of this study and possible future extensions, and (4) the frontiers of Network Ecology.

## 4.1 Forces Shaping Network Ecology

Our results show Network Ecology to be a broad and rapidly growing area of research. This finding is consistent with previous research both in ecology and in other areas of network science. Ings et al. (2009) used a more constrained survey of papers published in 12 selected journals to show the rapid increase of networks in ecology between 1970 and 2007. For 2007, they found that approximately 12% of the papers published in these journals were related to food webs, host-parasitod, or mutualist networks. This rapid expansion in ecology mirrors developments in other fields. For example while network modeling and analysis has a long history in the social sciences (Wasserman and Faust, 1994; Freeman, 2004), Borgatti and Foster (2003) document a similar exponential growth in social network publications between 1970 and 2000. Several drivers may explain the observed proliferation of Network Ecology and the notable discontinuous jump in publication rate between 1990 and 1991.

One possible factor driving the rise of Network Ecology and network science more generally is the creation of the Internet and the rise of the World Wide Web. Not only did the creation of computer networks generate a new science domain focused on network problems, these tools also broadened the general understanding of networks and the power of connectivity. As the commercial Internet become available in the late 1980s and early 1990s, this may have stimulated a broader scientific application of network techniques and crystallized the vocabulary of interactions and relational sciences around network and graph theoretical terms. This influence might be particularly important in explaining the observed jump in publications between 1990 and 1991.

A second possible driving factor could be that a more general network perspective had reached a critical mass of related knowledge and tools. As described already, the application and development of graph theory in general and network models and analysis in particular has a long history in several domains including ecology and sociology (Freeman, 2004; Wasserman and Faust, 1994; Fath and Patten, 1999; Newman et al., 2006). As network science pioneers had solved essential methodological problems and shown the usefulness of applying related concepts and tools, they might have began to attract additional scholars to a network perspective including those working within the area of ecology.

Related to this critical mass idea, a third possible explanation for the increase is the maturation of network modeling and analysis methodologies such that they could be applied more broadly. A popular explanation for the rapid increase in the social science literature in the late 1990s and 2000s is the entrance of physicist who applied their energy and quantitative machinery to sociological network problems. For example, Watts and Strogatz's (1998) paper on the small-world phenomenon is often cited as a tipping point in the development of network science. This publication was followed closely by Barabási and Albert's (1999) work examining the distribution of degree centralities in network models of several types of complex systems, and their subsequent calculation of the diameter of the World Wide Web (Albert et al., 1999). These papers generated much interest that led to both new mathematics and science that is surely part of the rise of Network Ecology. However, this does not adequately explain the Network Ecology publication acceleration between 1990 and 1991 as the jump preceded these publications.

A fourth potential factor in the rise of Network Ecology is our increasing ability to collect, store, and access data of all types (Stafford, 1993; Michener et al., 1998; Schmidt, 2003; Recknagel, 2006). Network models are data intensive. Thus, the data must be available for them to be a useful tool for empirical sciences. For example, the methodological innovation of using stable isotopes to estimate diet and trophic position in the food web (Peterson et al., 1985; Peterson and Fry, 1987) has enabled an increase in the development and quantification of food web data.

A fifth potential factor driving the rise of Network Ecology relates to the synthetic nature of constructing network models and their analysis. There appears to be a long-period cycle in science that swings between favoring more reductive scientific methods to favoring more synthetic or holistic approaches. Barabási (2012) argues that the rise of network science is in part because we are in a period of science in which we have reached the current limits of reductionism in many domains. Despite their typical presentation as a mutually exclusive dichotomous choice, it seems that science requires both reductionist and holistic approaches to be successful. As synthetic tools, network models provide a way to generate a more holistic understanding of a system by synthesizing data that might have been collected through more reductive methods.

In summary, all five of these factors have likely contributed to the rapid growth of network science in general and Network Ecology in particular. Fundamentally, ecology is a relational science; its central questions are about the relationships among species and their physical, chemical, biological, and social environments and how this ultimately creates and constrains the empirically observed patterns of species distribution, abundance, and evolution. Because network concepts, data representations, and analytical tools are broadly useful to address these relational questions, they have also become important within ecology forming Network Ecology as a subfield.

## 4.2 Networks, Systems, and Ecology

Following Borrett et al. (2012), we have defined Network Ecology by the use of a common concept (network) and related analytical methodologies rather than something more substantive such as common questions or focal empirical entities. This approach may prove problematic as a way to define a research area, and it may help explain the topic breadth and heterogeneity as well as the observed structure of co-authorship. Despite the possible inherent limitations of this definition, there may more commonality in this domain than is initially evident.

There is a strong if not always explicitly stated connection between networks and systems. This connection becomes most obvious by comparing the mathematical definitions of a network and a system. Klir (2001) provides a common sense definition of a system S as being comprised of a set of parts or things T and the relationship(s) R that connect them, S = (T, R). This parallels the definition of a graph G underlying any network model as being comprised of a set of vertices V and the edges E that connect them, G = (V, E). Thus, though the nomenclature and symbols are different, network and systems scientists appear to share the same fundamental research subject. Given the fundamental nature of systems ideas within ecology, the affinity between networks and systems and the commensurability of related approaches maybe one reason for the wide spread utility of network science. Network science, especially the quest to identify properties common across network types (Barabási, 2002), seems to be a subset of the more general systems science (von Bertalanffy, 1968; Klir, 2001, Patten et al., 2002).

Given this connection, it seems natural to claim that all of Network Ecology as defined here is a form of systems ecology. This would imply that systems ecology is a much broader subfield of ecology than we might otherwise expect. Indeed, we suspect that few of the authors of papers in our corpus self identify as systems ecologists. This is likely in part because published definitions of systems ecology assert that the ecosystem is the only valid object of study. For example, H.T. Odum's (1994: page ix) systems ecology textbook defined systems ecology as "the study of whole ecosystems and includes measurements of overall performance as well as a study of the details of systems design by which the overall behavior is produced from separate parts and mechanisms." Similarly, Jørgensen defined the field of systems ecology in his textbook by stating "it focuses on the properties of ecosystems and tries to reveal them by use of a systems approach" (Jørgensen, 2012: page 1). In doing so, Jørgensen directly equates systems ecology with ecosystem theory. In both cases the authors appear to be using a standard definition of an ecosystem as the study of the biological community with its non-living environment, which is conceptually a level of organization between communities and landscapes (but see Allen and Hoekstra, 1992).

Such definitions of systems ecology seem overly narrow. Ecologists can and do approach their subject matter at multiple levels of organization with a systems approach, be it a social community of Orca whales (Foster et al., 2012), placement of conservation areas (Rebelo and Siegfried, 1992; Sala et al., 2002), or estuarine ecosystems (Baird et al., 2004; Niquil et al., 2012). In this sense, a more encompassing definition of systems ecology was offered by Van Dyne (1966). While Van Dyne also defined systems ecology as the "study of the development, dynamics, and disruption of ecosystems", he uses the term ecosystem in a broader sense suggesting that we could apply the concept at the tissue level of organization as well as the more traditional ecosystem definition.

Given both (1) this broader understanding of systems ecology and (2) the fundamental affinity between a systems and a network approach, the documented rise and spread of Network Ecology becomes not only understandable, but perhaps even expected.

### 4.3 Limitations and Future Work

In this section we identify several limitations of our dataset and analysis and consider their potential impact on our findings. We conclude this section by considering a number of possible next steps for this research.

First, the scope of our bibliographic corpus is limited due to restricting our search to a subset of the ISI WoS. Thus, our search missed publications, topics, and scientific collaborations that might have emerged if we had expanded our search to conference proceedings or an additional database such as BIOSIS, MEDLINE, or Scopus. We selected WoS because of its strong and long-term coverage of ecological journals, where ecologists tend to publish their high quality work. Further, we used the subject or research area categories in WoS to help identify ecology publications. As these subject classifications are absent from BIOSIS and MEDLINE, we would have had to change our search method if we had included those databases. While the specific results would certainly be different, we do not expect that the broad trends identified would be substantively different if we used a different or combined set of databases.

A second limitation of our bibliographic corpus relates directly to our use of the WoS subject classifications as a means of identifying ecology publications. These research area classifications are applied to the whole journal, rather than being specific to an article. This helps explain why our corpus missed 36 (40%) of the Network Ecology articles between 2007 and 2012 identified by our experts. 21 of the articles missed (58%) were published in journals with a more general scope as indicted by their WoS research area classification of *Science & Technology – Other Topics*. For example, the articles by Allesina and Levine (2011) and Allesina and Tang (2012) are both clearly about Network Ecology. However, the first was published in the Proceedings of the National Academy of Sciences and the second was published in the journal Nature, both of which WoS labeled with a research area of *Science & Technology – Other Topics*. Thus, since the term ecology was not found in the title, abstract, or keywords, these papers were not identified as ecological articles in our search. Both articles were identifiable by our network search terms. This issue might also bias against selected Network Ecology authors who tend to publish their work in journals with different research area classifications. For example, of the 40 articles published by R.E. Ulanowicz through 2012 indexed by WoS with our *network* keywords, only 22 or 55% of these articles were published in a journal labeled as *Environmental Science & Ecology*. 9 or 23% of these publications were published in journals labeled by WoS as *Life Sciences Biomedical Other Topics*. This limitation suggests that our corpus provides a conservative estimate of the magnitude of Network Ecology.

A third limitation lies in the WoS database itself. Not all records in the WoS database are complete. Again, we can illustrate this with an example of a paper we would have expected to be in our corpus but that was not. Baird and Ulanowicz (1989) studied the seasonal dynamics of carbon flux in the Chesapeake Bay by building and analyzing network models to understand the ecosystem, and the article has been an influential Network Ecology paper as evidenced by its 397 cites as of this writing. It was published in Ecological Monographs, which has a research area classification of *Environmental Science & Ecology*, so it was identified as an ecology paper by our search. However, our Network Ecology search missed this paper because the information provided by the WoS record is missing the author keywords and abstract. This renders the WoS record invisible to our keyword term search. Inspection of the paper shows that both our search terms *network* and *web* appear in the abstract and keywords of the paper, so if this information had been included in the WoS database our search would have discovered this paper. As older WoS records tend to be less complete (personal observation), this issue does provide a systematic bias to our sampling that suggests our corpus underrepresents the earlier Network Ecology research.

Another challenge to our corpus works in an opposing direction – possibly inflating the volume of our results. In contrast to the Ings et al. (2009) study, we deliberately used a set of broad search terms to identify ecological science and the use of network concepts, techniques and tools within it. Despite the exclusion criteria we subsequently applied, our final corpus still included papers that did not strictly pertain to the area of Network Ecology we sought. Two papers exemplify this point. First, as described in section 3.1 the earliest record in our corpus is Griswold and Crowell (1936). This article was captured in our corpus due to its ecological science publication source and the term *webbing* in the title, which is part of the common name of the insect studied. However, this paper does not report the kind of network science we tried to capture. Second, Prodi and Thomas (1983) describe work based on an Italian network of sun-photometers. This work only uses the term *network* to refer to the sensor array. While this example is a common way of using the term *network*, it is distinct from our targeted research. Despite our term exclusions, it is likely that in a similar manner other papers were included in our corpus due to their use of *network* as referring to sensor and research networks in particular (e.g., Long Term Ecological Research Network, National Ecological Observatory Network). Although this will have inflated our corpus to some unknown degree, this problem is difficult to avoid and we decided to error on the side of being more inclusive.

We see two specific next steps to extend this research. First, while we have conducted a first analysis of the temporal dynamics of the topics in Network Ecology, we have not considered the temporal dynamics of the co-authorship structure. The current co-authorship network is cumulative, which may obfuscate the changing impact of particular individuals. Further, more experienced scientists will tend to be more central because they have had more time to publish and develop collaborations. Second, we could extend this analysis by developing a co-citation graph for the corpus. This network would provide a way to identify the commonality among papers suggested by their joint citations as well as identify influential publications along the lines of a traditional citation analysis.

## 4.4 Frontiers of Network Ecology

The future of Network Ecology appears bright. This is suggested by the rapid and sustained increase in Network Ecology publications (Fig. 1). Furthermore, we see at least four frontiers at which Network Ecology might develop further.

One frontier is the application of network concepts, techniques and tools to new areas within the broad field of ecology. For example, Cohen et al. (2012) proposed the development of physiological regulatory networks to investigate organismal ecology and evolution. Similarly, several scientists have begun to use ecosystem network analysis to investigate the sustainability of urban metabolisms (Zhang et al., 2010; Chen and Chen, 2012; Li et al., 2012) and industrial networks (Layton et al., 2012).

A second frontier is the application of existing Network Ecology to address applied questions. Memmott (2009) illustrates many possible ways in this regard. For example, Network Ecology can be used to assess the effectiveness of management and restoration or the potential impact of climate change. Accordingly, Hines et al. (2012, in review) use a comparative network approach to predict the potential impact of sea level rise on the sedimentary nitrogen cycle of the Cape Fear River estuary.

A third frontier emerges from attempts to apply Network Ecology more broadly. To do this effectively, Network Ecology will need to continue to develop methods for model construction (Fath et al., 2007) and overcome sampling and data limitations (Polis 1991). Linear inverse modeling is one tool to assist with this (Vézina and Pace, 1994; van Oevelen et al. 2010). Network Ecology will also need to improve its ability to quantify statistical uncertainty in network models and related implications for analysis and conclusions (Borrett and Osidele, 2007; Kones et al., 2009; Kaufman and Borrett, 2010).

A fourth frontier is the combination of multiple network perspectives and models. Fontaine et al. (2011) note that each network model provides a particular and necessarily limited perspective on the ecological systems being studied. Combining multiple perspectives has already led to new ecological insights. For example, Knight et al. (2005) combined both a food web network and a pollination network to show how a predatory fish can facilitate the fitness of nearby terrestrial plants. Likewise, Malcom (2011) applied an individual based genotype model to a network population model to show that the size and connectance pattern of genetic networks can change the trait heritability in the population and the population recovery from disturbance. In doing so, Malcom also illustrates the opportunity for crossing scales of analysis with network approaches.

These frontiers are only illustrative of the many possible directions that Network Ecology can grow into. Its future will be further enabled by developments in network science more generally. For example, statisticians are beginning to tackle the quantitative challenges of making inferences with networks (e.g., Kolaczyk, 2009) and the challenge to visualize the structural complexity that network approaches can capture (e.g., Moody et al., 2005; Lima, 2011).

## 4.5 Summary

Network Ecology is comprised of scientists using network models to investigate ecological systems at many different hierarchical levels of organization. Network Ecology is defined by the use of a general model – a network – and an analysis that resembles an instance of a general systems approach. It is a large and rapidly growing subfield of ecology. As such, it represents one approach in a broadly defined systems ecology, which has many types of specific ecological instantiations. This is reflected in the broad set of topics identified in our Network Ecology corpus and the large and fragmented co-authorship structure of these publications. However, these scientists broadly use network concepts, techniques, and tools to (1) characterize the system organization (Ulanowicz, 1983; Croft et al., 2004; Borrett, 2013), (2) investigate the consequences of the network organization (Dunne et al., 2002; Borrett et al., 2006; Allesina and Pascual, 2009), and (3) identify the processes or mechanisms that might generate the observed patterns (Ulanowicz, 1986; Williams and Martinez, 2000; Fath and Killian, 2007; Guimãres et al., 2007; Allesina et al., 2008). Network Ecology is a research area with a long history and bright future in part because as Lima (2011) said “Networks are everywhere”.

# 5 Acknowledgements

This work was initially presented at the “Systems Ecology: A Network Perspective and Retrospective” symposium in honor of Dr. Bernard C. Patten’s work at the University of Georgia (April, 2012). The work benefited from critiques by David Hines, Emily Oxe, Brian Fath and two anonymous reviewers. We also gratefully acknowledge support for this work by the UNCW Charles L. Cahill Award (SRB) and by the US National Science Foundation (DEB1020944).

# 7 List of Tables and Figures with captions

Table 1: Percent of ecology publications that include selected concepts and methods in ecology, including the 50 most important concepts in ecology as ranked by British ecologists in 1986 (Cherrett, 1989).

Table 2. Number of records returned using search terms for two dimensions of Network Ecology: Ecological science (ISI research areas *Environmental Science & Ecology* and *Evolutionary Biology* and topic *ecology*) and network science (topics *network*, *graph theory* and *web*).

Table 3. Most frequently cited papers by experts asked to identify important Network Ecology papers published between 2007 and 2012.

Table 4. Topic cluster distributions and descriptions of the largest 5 topic clusters in the whole network and in the six temporal subsets.

Table 5. Three authors with highest betweenness centrality in the largest co-authorship clusters.

Figure 1: (a) Number of Network Ecology publications per year, and (b) as percent of total ecology publications discovered in our bibliographic search.

Figure 2: Top 20 ISI research areas (a) and journal titles (b) for Network Ecology research.

Figure 3: Citation distribution for articles in the Network Ecology corpus.

Figure 4: Topic network construction example: (a) the one step local neighborhood around the exemplar paper Burgmer et al. (2010) which is colored blue, and (b) the topic clusters that appear 3 steps from the focal paper. A contour map overlay created with a 2-d kernel density surface algorithm also shows the paper clusters in (b). Clusters are labeled with the most common terms.

Figure 5: Topic network contour plot. Regions are labeled with the most common terms found in the clusters and font size corresponds to term frequency.

Figure 6. Frequency distribution of the number of co-authors of Network Ecology articles and the temporal trend in collaboration (inset).

Figure 7. Collaboration by scientists publishing in Network Ecology as indicted by co-authorship (network node size and color proportional to degree, contour lines capture overall density of the academic field).

Figure 8. Clustering in co-authorship network for Network Ecology publications. Each panel highlights distinct clusters in the co-authorship network and indicates the cluster size, dispersion, and the most central author (betweenness).

Figure 9: One-step ego-networks of (a) Bernard C. Patten and R.E. Ulanowicz, (b) S.R. Carpenter and (c) D.C.G. Muir.

# 8 Tables

**Table 1:** Percent of ecology publications that include selected concepts and methods in ecology, including the 50 most important concepts in ecology as ranked by British ecologists in 1986 (Cherrett, 1989).

| 2012 Rank | Concept[#] | N | % of 64,115 | 1986 Rank |
|---|---|---|---|---|
| 1 | species | 17,126 | 26.7% | |
| 2 | model | 16,853 | 26.3% | |
| 3 | system | 12,707 | 19.8% | |
| 4 | population cycles | 11,225 | 17.5% | 19 |
| 5 | pattern | 9,158 | 14.3% | 32 |
| 6 | community | 8,456 | 13.2% | 8 |
| 7 | ecosystem | 6,853 | 10.7% | 1 |
| 8 | species diversity | 6,448 | 10.1% | 14 |
| 9 | evolution | 5,384 | 8.4% | |
| 10 | conservation of resources | 4,926 | 7.7% | 4 |
| 11 | energy flow | 4,897 | 7.6% | 3 |
| 12 | materials cycling | 3,808 | 5.9% | 7 |
| 13 | landscape | 3,387 | 5.3% | |
| 14 | indicator organisms | 2,698 | 4.2% | 29 |
| 15 | organism | 2,680 | 4.2% | |
| 16 | regression | 2,337 | 3.6% | |
| 17 | ecological adaptation | 2,200 | 3.4% | 12 |
| 18 | disease | 2,136 | 3.3% | |
| 19 | competition | 1,939 | 3.0% | 5 |
| 20 | natural disturbance | 1,661 | 2.6% | 26 |
| 21 | systems ecology | 1,634 | 2.5% | |
| 22 | trophic level | 1,415 | 2.2% | 31 |
| 23 | habitat restoration | 1,344 | 2.1% | 27 |
| 24 | limiting factors | 1,220 | 1.9% | 16 |
| 25 | niche | 1,108 | 1.7% | 6 |
| 26 | food webs | 1,017 | 1.6% | 11 |
| 27 | stable isotope | 882 | 1.4% | |
| 28 | predator-prey interactions | 790 | 1.2% | 20 |
| 29 | anova or analysis of variance | 758 | 1.2% | |
| 30 | species-area relationships | 751 | 1.2% | 39 |
| 31 | bioaccumulation in food chains | 704 | 1.1% | 23 |
| 32 | succession | 595 | 0.9% | 2 |
| 33 | plant herbivore | 555 | 0.9% | 21 |
| 34 | 3/2 thinning law | 551 | 0.9% | 49 |
| 35 | managed reserve | 466 | 0.7% | 28 |
| 36 | density dependent regulation | 424 | 0.7% | 15 |
| 37 | environmental heterogeneity | 394 | 0.6% | 13 |
| 38 | life history strategies | 370 | 0.6% | 9 |
| 39 | carrying capacity | 363 | 0.6% | 17 |
| 40 | coevolution | 263 | 0.4% | 24 |
| 41 | biome | 252 | 0.4% | 47 |
| 42 | (diversity or biodiversity) and stability | 248 | 0.4% | 35 |
| 43 | guild | 240 | 0.4% | 50 |
| 44 | stochastic processes | 208 | 0.3% | 25 |
| 45 | island biogeography or biogeographic theory | 178 | 0.3% | 22 |
| 46 | allometry | 174 | 0.3% | |
| 47 | parasite-host interactions | 170 | 0.3% | 38 |
| 48 | structural equation modeling | 131 | 0.2% | |
| 49 | ecotype | 105 | 0.2% | 40 |
| 50 | species packing | 87 | 0.1% | 48 |
| 51 | keystone species | 85 | 0.1% | 46 |
| 52 | allocation theory | 78 | 0.1% | 43 |
| 53 | optimal foraging | 72 | 0.1% | 37 |
| 54 | competition and species exclusion | 63 | 0.1% | 30 |
| 55 | territoriality | 58 | 0.1% | 42 |
| 56 | maximum sustainable yield | 32 | 0.0% | 18 |
| 57 | climax | 25 | 0.0% | 41 |
| 58 | pyramid of numbers | 14 | 0.0% | 45 |
| 59 | r and K selection | 11 | 0.0% | 33 |
| 60 | intrinsic regulation | 9 | 0.0% | 44 |
| 61 | socioecology | 8 | 0.0% | 36 |
| 62 | plant-animal coevolution | 7 | 0.0% | 34 |
| 63 | ecosystem fragility | 5 | 0.0% | 10 |

**Note:** Plural forms of concepts listed were also searched and are included in the counts. For concepts and methods that seemed overly specific, we selected to search a more general form (excluding greyed out words). For example, the phrase *energy flow* is very specific, so we also searched the term *energy* by itself. We added the 13 concepts that are not ranked in the BES 1986 survey.

**Table 2.** Number of records returned using search terms for two dimensions of Network Ecology: Ecological science (ISI research areas *Environmental Science & Ecology* and *Evolutionary Biology* and topic *ecology*) and network science (topics *network*, *graph theory* and *web*).

| | | | Ecology Dimension | | | | | |
|---|---|---|---|---|---|---|---|---|
| | | | **null** | **Environmental Science & Ecology** | **Evolutionary Biology** | **Ecology** | | |
| | | | **0** | **A** | **B** | **C** | **A or B** | **A or B or C** |
| **Network Dimension** | **null** | **0** | | 1,002,490 | 139,706 | 106,825 | 1,075,483 | 1,144,552 |
| | **Network** | **X** | 666,367 | 19,987 | 1,900 | 2,751 | 21,005 | 22,534 |
| | **Web** | **Y** | 110,868 | 10,567 | 846 | 2,978 | 10,749 | 12,196 |
| | **Graph Theory** | **Z** | 6,198 | 183 | 22 | 76 | 189 | 207 |
| | | **X or Y or Z** | 769,435 | 29,877 | 2,688 | 5,450 | 31,068 | **33,900** |

**Note:** Number at the intersection of the search terms are the set intersection (∩) of the row and column terms; "or" in the table represents the set union (∪) of search terms.

**Table 3.** Most frequently cited papers by experts asked to identify important Network Ecology papers published between 2007 and 2012.

| Authors | Year | Title | Journal |
|---|---|---|---|
| **Mentioned Three Times** | | | |
| Dale and Fortin | 2010 | From graphs to spatial graphs | Ann. Rev. Ecol. Evol. Syst. |
| Fontaine et al. | 2011 | The ecological and evolutionary implications of merging different types of networks | Ecol. Lett. |
| **Mentioned Two Times** | | | |
| Aizen et al. | 2012 | Specialization and Rarity Predict Nonrandom Loss of Interactions from Mutualist Networks. | Science |
| Allesina and Pascual | 2009 | Googling Food Webs: Can an Eigenvector Measure Species' Importance for Coextinctions? | PLoS Comp. Bio. |
| Baird et al. | 2008 | Nutrient dynamics in the Sylt-Romo Bight ecosystem, German Wadden Sea: An ecological network analysis approach | Estuar. Coast. Shelf Sci. |
| Bascompte and Jordano | 2007 | Plant-animal mutualistic networks: The architecture of biodiversity | Ann. Rev. Ecol. Evol. Syst. |
| Berlow et al. | 2009 | Simple prediction of interaction strengths in complex food webs | Proc. Nat. Acad. Sci. USA |
| Chen et al | 2008 | Network position of hosts in food webs and their parasite diversity | Oikos |
| Dunne and Williams | 2009 | Cascading extinctions and community collapse in model food webs | Philos. Trans. R. Soc. Lond. B |
| Nuismer et al. | 2013 | Coevolution and the architecture of mutualistic networks | Evolution |
| Otto et al. | 2007 | Allometric degree distributions facilitate food-web stability | Nature |
| Ramirez | 2012 | Poplation persistence under advection-diffusion in river networks | J. Math. Biol. |
| Thébault and Fontaine | 2010 | Stability of ecologcial communities and the architecture of mutualistic and trophic networks | Science |
| Urban et al. | 2009 | Graph models of habitat mosaics | Ecol. Lett. |
| Wey et al. | 2008 | Social network analysis of animal behaviour: a promising tool for the study of sociality | Animal Behavior |

**Table 4.** Topic cluster distributions and descriptions of the largest 5 topic clusters in the whole network and in the six temporal subsets.

| | | |
|---|---|---|
| *Total: P=21,636, C=86, M=0.927, TH =0.98, Mean Cluster Size (std): 251 (185)* | | |
| *Rank* | *Prop* | *Most Common Terms* |
| 1 | 0.03781 | conservation, area, forest, species |
| 2 | 0.03360 | lake, fish, web, concentration |
| 3 | 0.03263 | phytoplankton, web, community, rate |
| 4 | 0.02935 | delta, isotope, stable, carbon |
| 5 | 0.02921 | predator, prey, web, community |
| *2010-2012: P=8606, C=37, M=0.828, TH =0.96, Mean Cluster Size (std): 232 (142)* | | |
| 1 | 0.06100 | conservation, habitat, landscape, area |
| 2 | 0.05415 | urban, city, network, system |
| 3 | 0.05403 | lake, community, temperature, web |
| 4 | 0.04892 | stream, river, water, flow |
| 5 | 0.04741 | species, interaction, plant, specie |
| *2005-2009: P=9764,C=41, M=0.824, TH =0.96, Mean Cluster Size (std): 238 (165)* | | |
| 1 | 0.06616 | network, development, innovation, paper |
| 2 | 0.05879 | conservation, area, forest, species |
| 3 | 0.05602 | stream, river, water, model |
| 4 | 0.05131 | water, system, model, result |
| 5 | 0.04998 | predator, prey, interaction, plant |
| *2000–2004: P=4828, C=31, M=0.831, TH =0.96, Mean Cluster Size (std): 156 (98)* | | |
| 1 | 0.07809 | reserve, conservation, specie, area |
| 2 | 0.07809 | predator, prey, community, specie |
| 3 | 0.06048 | innovation, regional, network, firm |
| 4 | 0.05634 | air, concentration, road, emission |
| 5 | 0.05282 | water, river, model, rainfall |
| *1995–1999: P=2665, C=31, M=0.838, TH =0.95, Mean Cluster Size (std): 86 (68))* | | |
| 1 | 0.09343 | network, development, firm, paper |
| 2 | 0.09231 | production, phytoplankton, bacterial, rate |
| 3 | 0.06942 | lake, fish, web, concentration |
| 4 | 0.06492 | conservation, area, forest, landscape |
| 5 | 0.06079 | specie, predator, interaction, prey |
| *1990–1994: P=1124,C=29, M=0.851, TH =0.95, Mean Cluster Size (std): 39 (25)* | | |
| 1 | 0.07473 | deposition, concentration, precipitation, urban |
| 2 | 0.07028 | web, specie, species, ecosystem |
| 3 | 0.07028 | region, development, information, network |
| 4 | 0.06139 | level, biomass, increase, production |
| 5 | 0.06139 | fish, isotope, specie, winter |
| *1980–1989: P=418, C=11, M=0.642, TH =0.90, Mean Cluster Size (std): 38 (15)* | | |
| 1 | 0.15311 | spider, web, prey, araneae |
| 2 | 0.14115 | distribution, model, national, water |
| 3 | 0.10766 | network, fracture, measure, simulation |
| 4 | 0.10766 | analysis, plan, relationship, network |
| 5 | 0.10048 | web, dynamic, pattern, landscape |

*P = papers in largest component, C =number of topic clusters, M = modularity score, TH =Topic Heterogeneity { = 1-sum($p_k^2$) }*

**Note:** A lower edge threshold is used for the individual waves (0.25) than for the total corpus (0.35) to counterbalance the lower power in judging paper similarity based on co-word frequency.

**Table 5.** Three authors with highest betweenness centrality in the largest co-authorship clusters.

| Cluster Size | Cluster ID | Most Central Authors | Weighted Degree Centrality | Number of Coauthors | Betweenness Centrality | Closeness Centrality |
|---|---|---|---|---|---|---|
| 1618 | 13 | Woodward, G. | 107 | 112 | 0.234 | 0.304 |
| | | Memmott, J. | 71 | 86 | 0.150 | 0.286 |
| | | Martinez, N. D. | 59 | 71 | 0.125 | 0.291 |
| 1603 | 11 | Muir, D. C. G. | 192 | 201 | 0.544 | 0.369 |
| | | Fisk, A. T. | 107 | 120 | 0.131 | 0.347 |
| | | Backus, S. M. | 52 | 60 | 0.126 | 0.330 |
| 1129 | 8 | Deruiter, P. C. | 75 | 88 | 0.373 | 0.316 |
| | | Scheu, S. | 79 | 97 | 0.183 | 0.268 |
| | | van der Putten, W. H. | 75 | 87 | 0.152 | 0.300 |
| 889 | 23 | McDowell, W. H. | 42 | 59 | 0.185 | 0.245 |
| | | Wollheim, W. M. | 48 | 59 | 0.160 | 0.266 |
| | | Hamilton.Stephen.K. | 24 | 34 | 0.159 | 0.246 |
| 798 | 17 | Vesala, T. | 206 | 236 | 0.279 | 0.450 |
| | | Hollinger.David.Y. | 113 | 144 | 0.115 | 0.415 |
| | | Wofsy, S. C. | 71 | 96 | 0.107 | 0.351 |
| 765 | 27 | Estes, J. A. | 46 | 55 | 0.281 | 0.159 |
| | | Demeester, L. | 29 | 39 | 0.246 | 0.142 |
| | | Guimãres, P. R. .Jr. | 42 | 51 | 0.238 | 0.155 |
| 765 | 42 | Possingham, H.P. | 80 | 89 | 0.303 | 0.263 |
| | | Gaines, S. D. | 38 | 50 | 0.286 | 0.269 |
| | | Halpern, B. S. | 39 | 47 | 0.193 | 0.279 |
| 729 | 51 | Kitchell, J. F. | 68 | 72 | 0.521 | 0.308 |
| | | Carpenter, S. R. | 72 | 73 | 0.361 | 0.295 |
| | | Stenseth, N. C. | 46 | 59 | 0.173 | 0.233 |
| 662 | 20 | Pressey, R. L. | 53 | 71 | 0.311 | 0.276 |
| | | Burgess, N. D. | 60 | 76 | 0.271 | 0.257 |
| | | Gaston. K. J. | 43 | 52 | 0.214 | 0.255 |
| 646 | 35 | Evers, D. C. | 54 | 72 | 0.412 | 0.248 |
| | | Murphy, C. A. | 13 | 19 | 0.183 | 0.207 |
| | | Gay, D. A. | 15 | 21 | 0.163 | 0.208 |
| 602 | 40 | Moilanen, A. | 52 | 65 | 0.635 | 0.265 |
| | | Hanski, I. | 23 | 24 | 0.261 | 0.238 |
| | | Zobe, .M. | 47 | 66 | 0.214 | 0.242 |
| 567 | 9 | Sime-Ngando, T. | 55 | 68 | 0.307 | 0.228 |
| | | Thingstad, T. F. | 41 | 54 | 0.278 | 0.245 |
| | | Nielsen, T. G. | 41 | 50 | 0.232 | 0.223 |
| 536 | 41 | Gessner, M. O. | 50 | 66 | 0.316 | 0.265 |
| | | Prieur-Richard, A. | 24 | 35 | 0.250 | 0.251 |
| | | Jongman, R. H. | 25 | 34 | 0.213 | 0.183 |
| 492 | 38 | Duarte, C. M. | 40 | 47 | 0.624 | 0.234 |
| | | Gasol, J. M. | 31 | 38 | 0.493 | 0.226 |
| | | Orth, R. J. W. | 23 | 30 | 0.195 | 0.215 |
| 486 | 36 | Scheffer, M. | 27 | 33 | 0.361 | 0.280 |
| | | Folke, C. S. | 50 | 63 | 0.290 | 0.272 |
| | | Mooij, W. M. | 35 | 45 | 0.243 | 0.250 |
| 437 | 29 | Tanabe, S. | 48 | 54 | 0.569 | 0.268 |
| | | Okuda, N. | 22 | 25 | 0.410 | 0.260 |
| | | Kohzu, A. | 31 | 39 | 0.301 | 0.234 |
| 419 | 74 | Hillebrand, H. | 29 | 30 | 0.360 | 0.264 |
| | | Cardinale, B. J. | 20 | 28 | 0.220 | 0.239 |
| | | Sommer, U. | 17 | 21 | 0.200 | 0.220 |
| 416 | 12 | Bathmann, U. | 66 | 95 | 0.444 | 0.316 |
| | | Wishner, K. F. | 42 | 60 | 0.269 | 0.279 |
| | | Wong, C. S. | 50 | 69 | 0.119 | 0.265 |
| 412 | 61 | Fagan, W. F. | 27 | 30 | 0.537 | 0.281 |
| | | Rodriguez-Iturbe, I. | 46 | 36 | 0.428 | 0.294 |
| | | Rinaldo, A. | 57 | 49 | 0.419 | 0.290 |
| 399 | 19 | Niquil, N. | 30 | 37 | 0.485 | 0.285 |
| | | Dupuy, C. | 50 | 64 | 0.438 | 0.274 |
| | | Marques, J. C. | 13 | 13 | 0.294 | 0.262 |
| 396 | 30 | Gremillet, D. | 37 | 45 | 0.482 | 0.291 |
| | | Cherel, Y. | 30 | 37 | 0.464 | 0.269 |
| | | Phillips, R. A. | 20 | 23 | 0.328 | 0.276 |
| 395 | 56 | Poulin, M. | 39 | 50 | 0.498 | 0.274 |
| | | Tremblay, J. | 28 | 36 | 0.327 | 0.270 |
| | | Lovejoy, C. | 21 | 26 | 0.271 | 0.253 |
| 391 | 88 | Edgerton, E. S. | 39 | 46 | 0.564 | 0.244 |
| | | Chow, J. C. | 76 | 93 | 0.447 | 0.229 |
| | | Fu, J. | 17 | 23 | 0.232 | 0.198 |
| 389 | 7 | Rabalais, N. N. | 32 | 45 | 0.442 | 0.250 |
| | | Kemp, W. M. | 27 | 38 | 0.253 | 0.228 |
| | | Gilbert, D. | 15 | 22 | 0.221 | 0.226 |
| 376 | 69 | Jochem, F. J. | 17 | 22 | 0.561 | 0.209 |
| | | van der Klift, M. A. | 25 | 26 | 0.522 | 0.202 |
| | | Peterson, B. J. | 15 | 17 | 0.499 | 0.206 |
| 374 | 73 | McGlynn, B. L. | 14 | 15 | 0.484 | 0.199 |
| | | Wondzell, S. M. | 10 | 11 | 0.396 | 0.184 |
| | | Swanson, F. J. | 13 | 17 | 0.387 | 0.174 |
| 366 | 3 | Davies, J. | 28 | 39 | 0.317 | 0.234 |
| | | Claude, H. | 28 | 41 | 0.216 | 0.256 |
| | | McAllister, R. R. J. | 19 | 26 | 0.207 | 0.204 |
| 365 | 84 | Patten, B. C. | 38 | 35 | 0.529 | 0.228 |
| | | Schramski, J. R. | 31 | 33 | 0.382 | 0.225 |
| | | Calbet, A. | 21 | 26 | 0.309 | 0.176 |
| 356 | 75 | Symondson, W. O. C. | 28 | 33 | 0.615 | 0.174 |
| | | Cruaud, C. | 20 | 28 | 0.551 | 0.180 |
| | | Olayemi, A. | 8 | 12 | 0.422 | 0.164 |
| 355 | 37 | Swenson, N. G. | 20 | 28 | 0.456 | 0.203 |
| | | Legendre, P. | 35 | 50 | 0.406 | 0.166 |
| | | Savage, V. M. | 20 | 23 | 0.404 | 0.203 |
| 354 | 33 | Ashley, R.M. | 29 | 39 | 0.515 | 0.213 |
| | | Vollertsen, J. | 24 | 25 | 0.412 | 0.209 |
| | | Yang, W. | 15 | 19 | 0.276 | 0.195 |
| 338 | 43 | Pierce, G. J. | 23 | 32 | 0.597 | 0.223 |
| | | Pond, D. W. | 19 | 24 | 0.512 | 0.226 |
| | | Fielding, S. | 15 | 21 | 0.232 | 0.195 |
| 330 | 1 | Lehtonen, K. K. | 26 | 37 | 0.461 | 0.195 |
| | | Vuorinen, P. J. | 14 | 19 | 0.440 | 0.179 |
| | | Hjorth, M. | 30 | 42 | 0.403 | 0.198 |
| 330 | 47 | Solidoro, C. | 16 | 22 | 0.559 | 0.228 |
| | | Park, Y. | 26 | 35 | 0.380 | 0.215 |
| | | Libralato, S. | 10 | 13 | 0.361 | 0.224 |
| 319 | 83 | Croft, D. P. | 30 | 35 | 0.440 | 0.218 |
| | | Codling, E. A. | 8 | 11 | 0.404 | 0.195 |
| | | James, R. | 32 | 37 | 0.399 | 0.215 |
| 318 | 10 | Blackledge, T. A. | 18 | 19 | 0.455 | 0.288 |
| | | Herberstein, M. E. | 31 | 32 | 0.361 | 0.250 |
| | | Coddington, J. A. | 19 | 23 | 0.319 | 0.279 |
| 313 | 39 | Moran, M. D. | 51 | 56 | 0.358 | 0.302 |
| | | Prank, M. | 54 | 61 | 0.266 | 0.293 |
| | | Christensen, J. H. | 46 | 50 | 0.239 | 0.294 |
| 301 | 21 | Tranvik, L. J. | 26 | 29 | 0.439 | 0.272 |
| | | Persson, L. | 34 | 35 | 0.410 | 0.264 |
| | | Nystrom, P. | 31 | 38 | 0.388 | 0.236 |

# 9 Figures

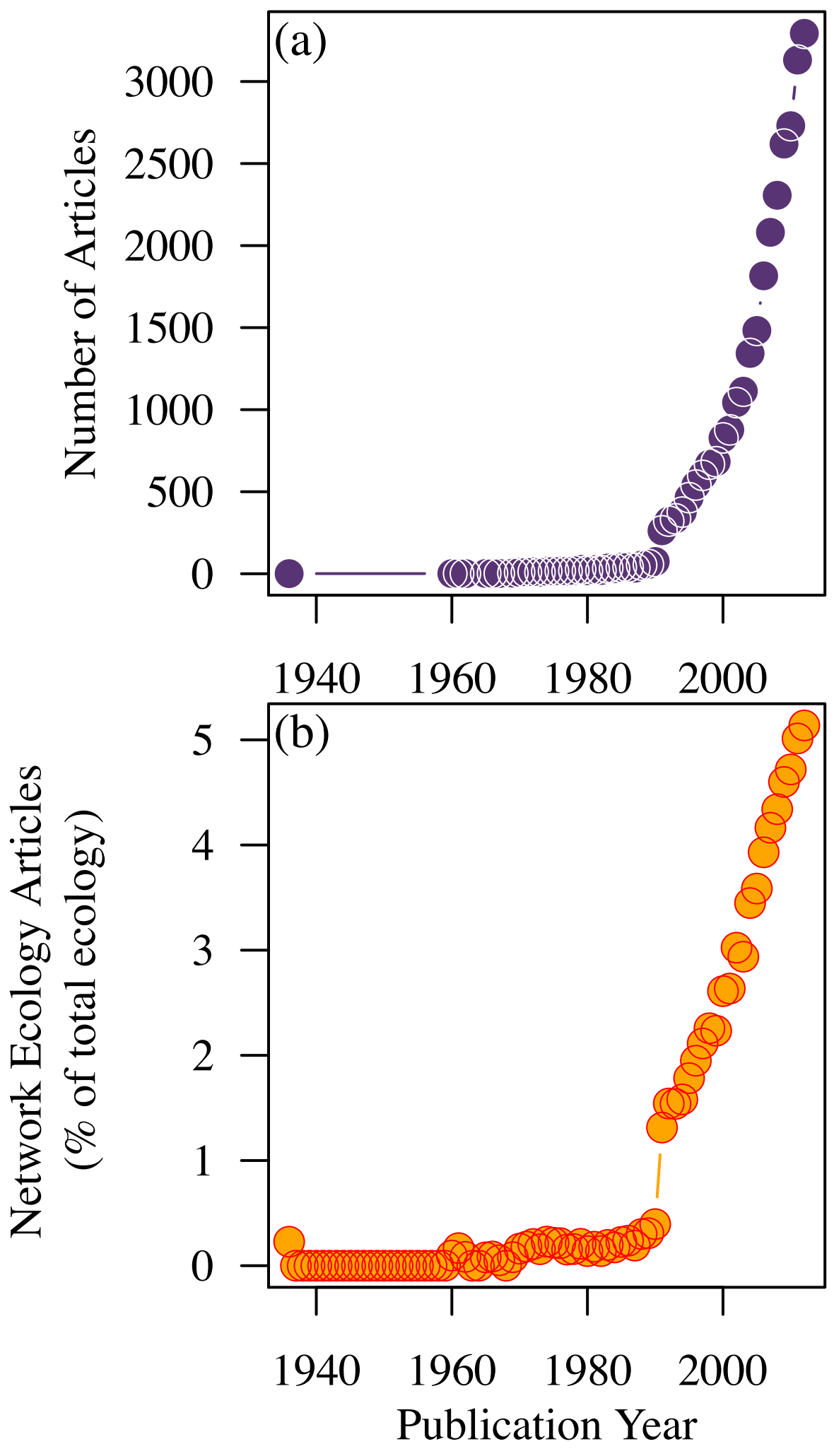

**Figure 1:** (a) Number of Network Ecology publications per year, and (b) as percent of total ecology publications discovered in our bibliographic search.

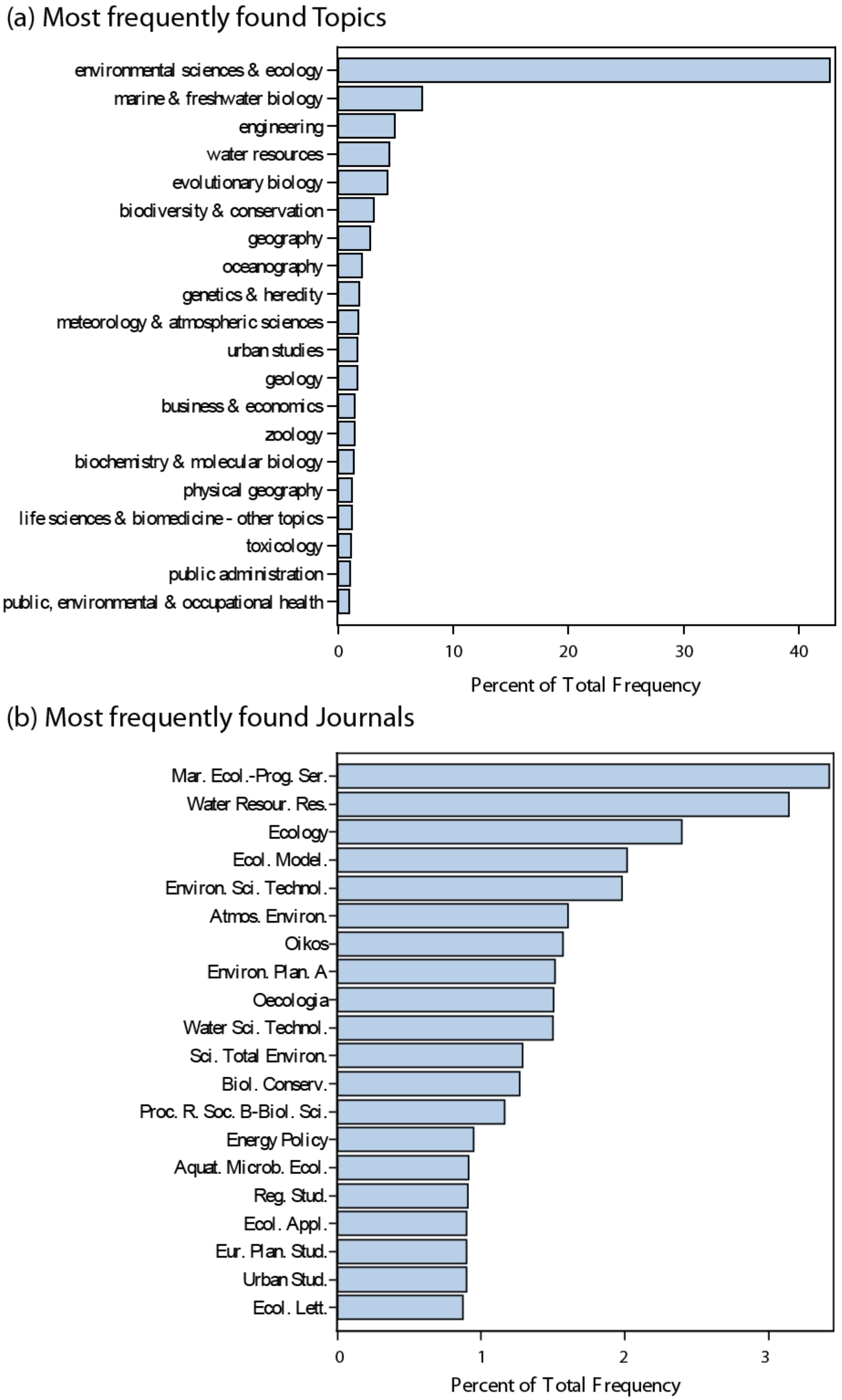

**Figure 2:** Top 20 ISI research areas (a) and journal titles (b) for Network Ecology research.

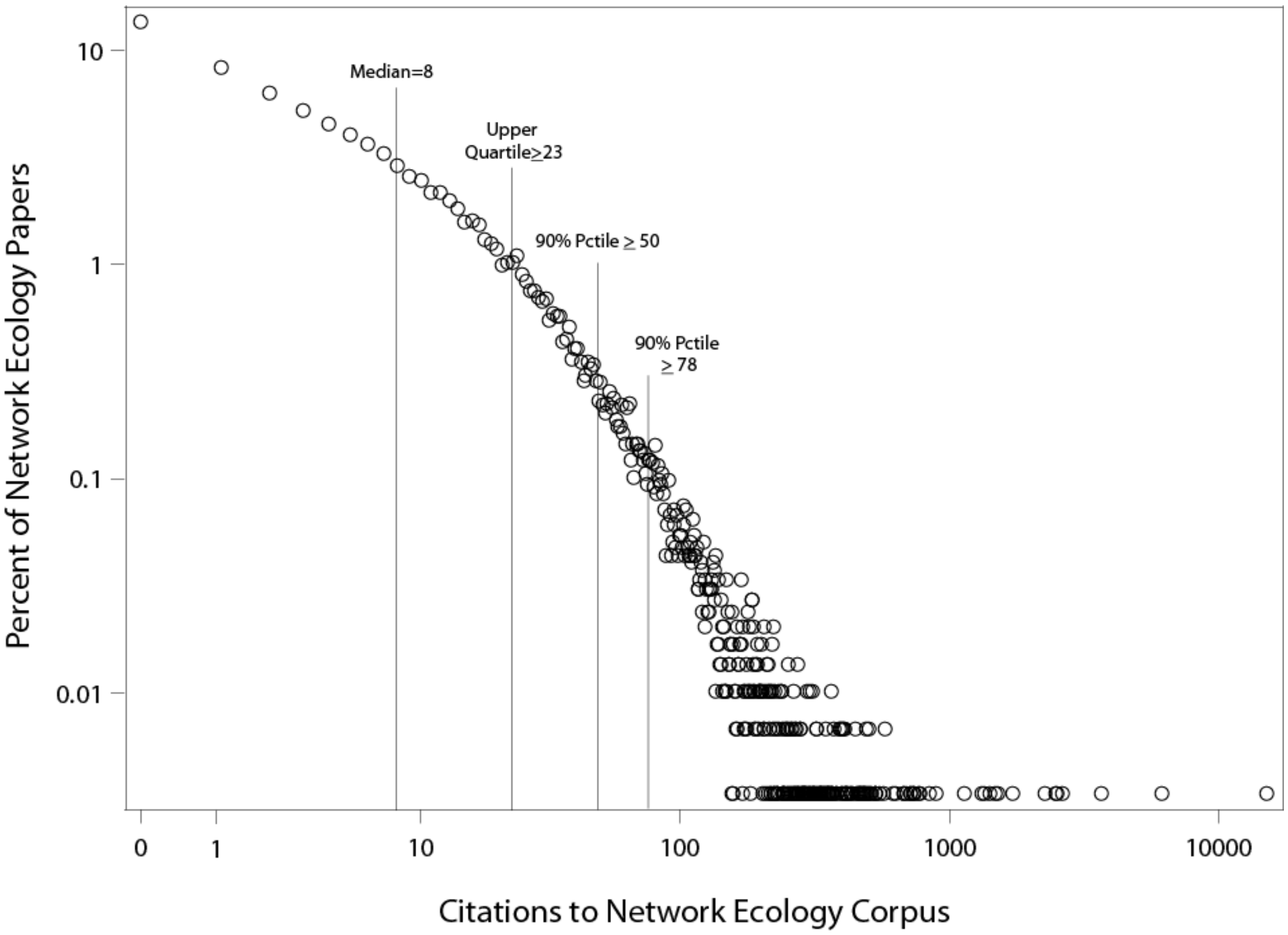

**Figure 3:** Citation distribution for articles in the Network Ecology corpus.

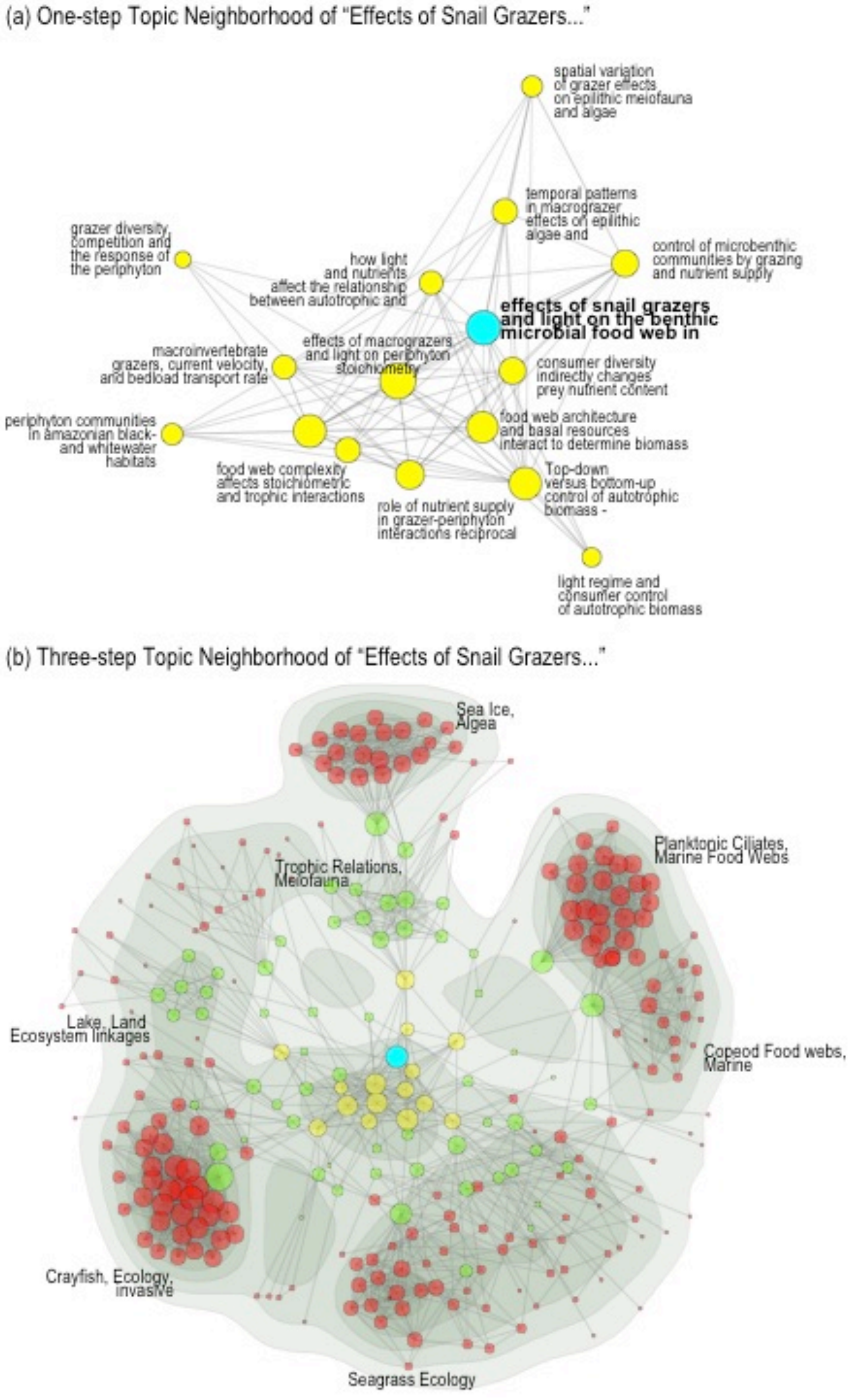

**Figure 4:** Topic network construction example: (a) the one step local neighborhood around the exemplar paper by Burgmer et al. (2010) which is colored blue, and (b) the topic clusters that appear 3 steps from the focal paper. A contour map overlay created with a 2-d kernel density surface algorithm also shows the paper clusters in (b). Clusters are labeled with the most common terms.

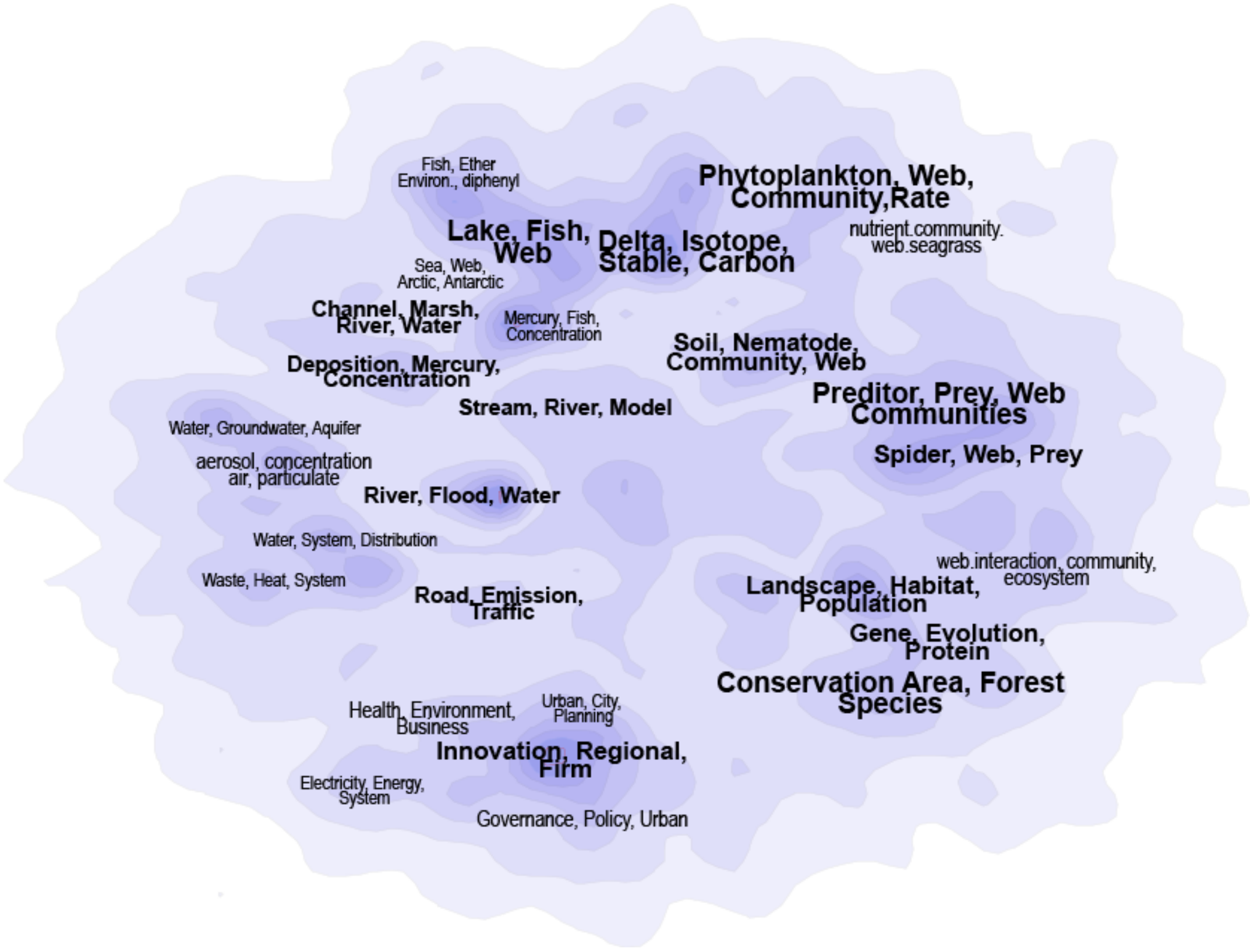

**Figure 5:** Topic network contour plot. Regions are labeled with the most common terms found in the clusters and font size corresponds to term frequency.

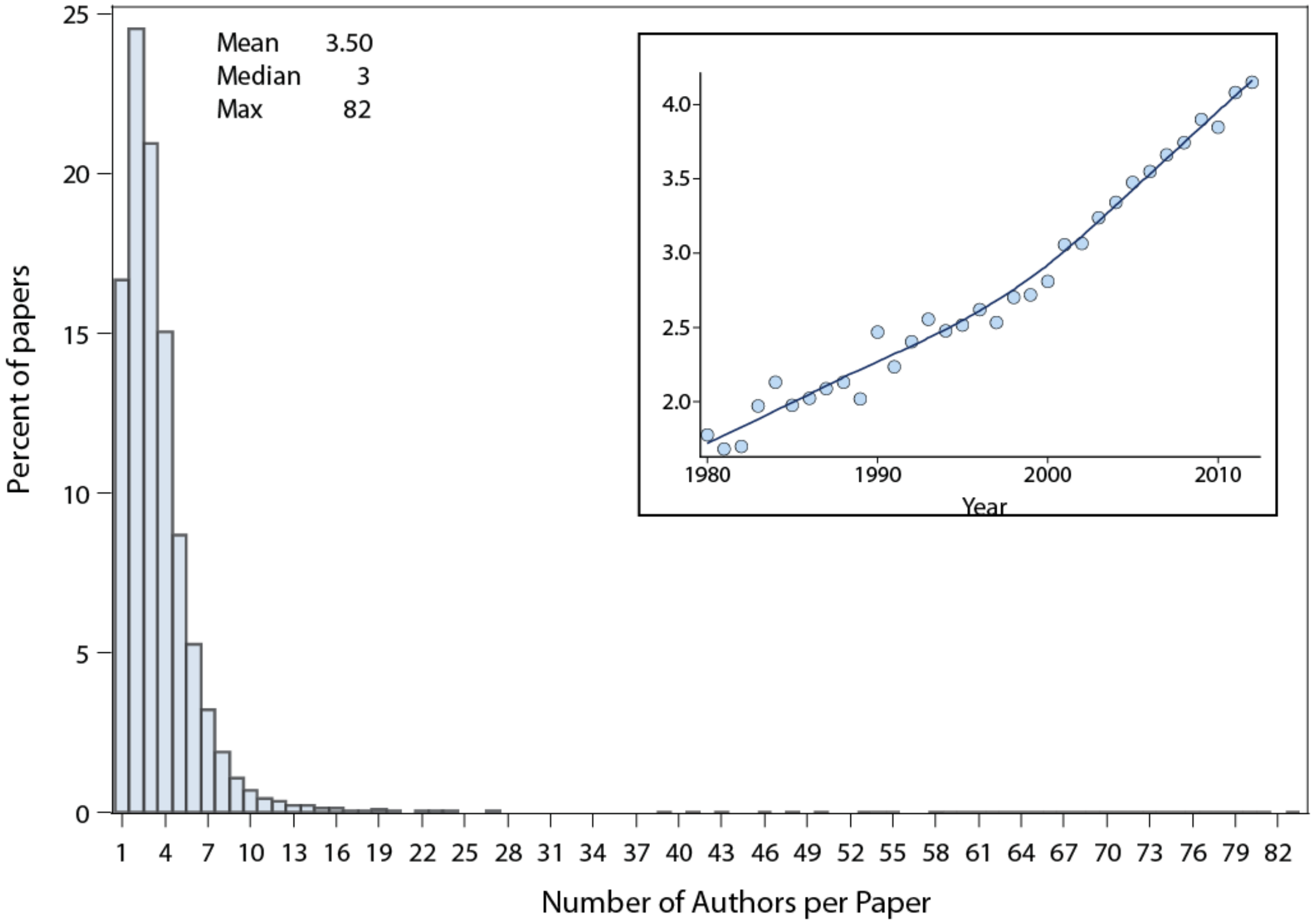

**Figure 6.** Frequency distribution of the number of co-authors of Network Ecology articles and the temporal trend in collaboration (inset).

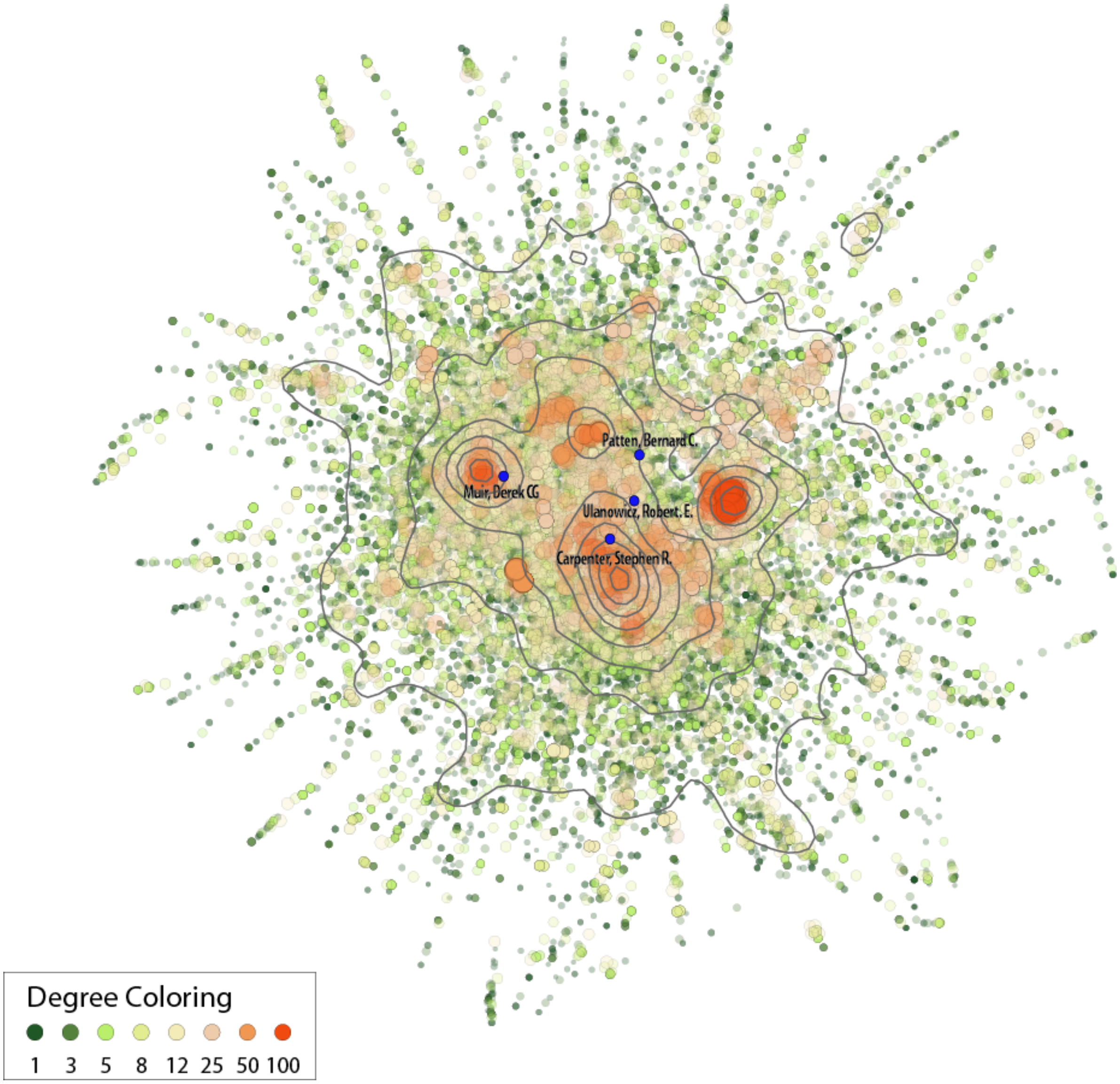

**Figure 7.** Collaboration by scientists publishing in Network Ecology as indicted by co-authorship (network node size and color proportional to degree, contour lines capture overall density of the academic field).

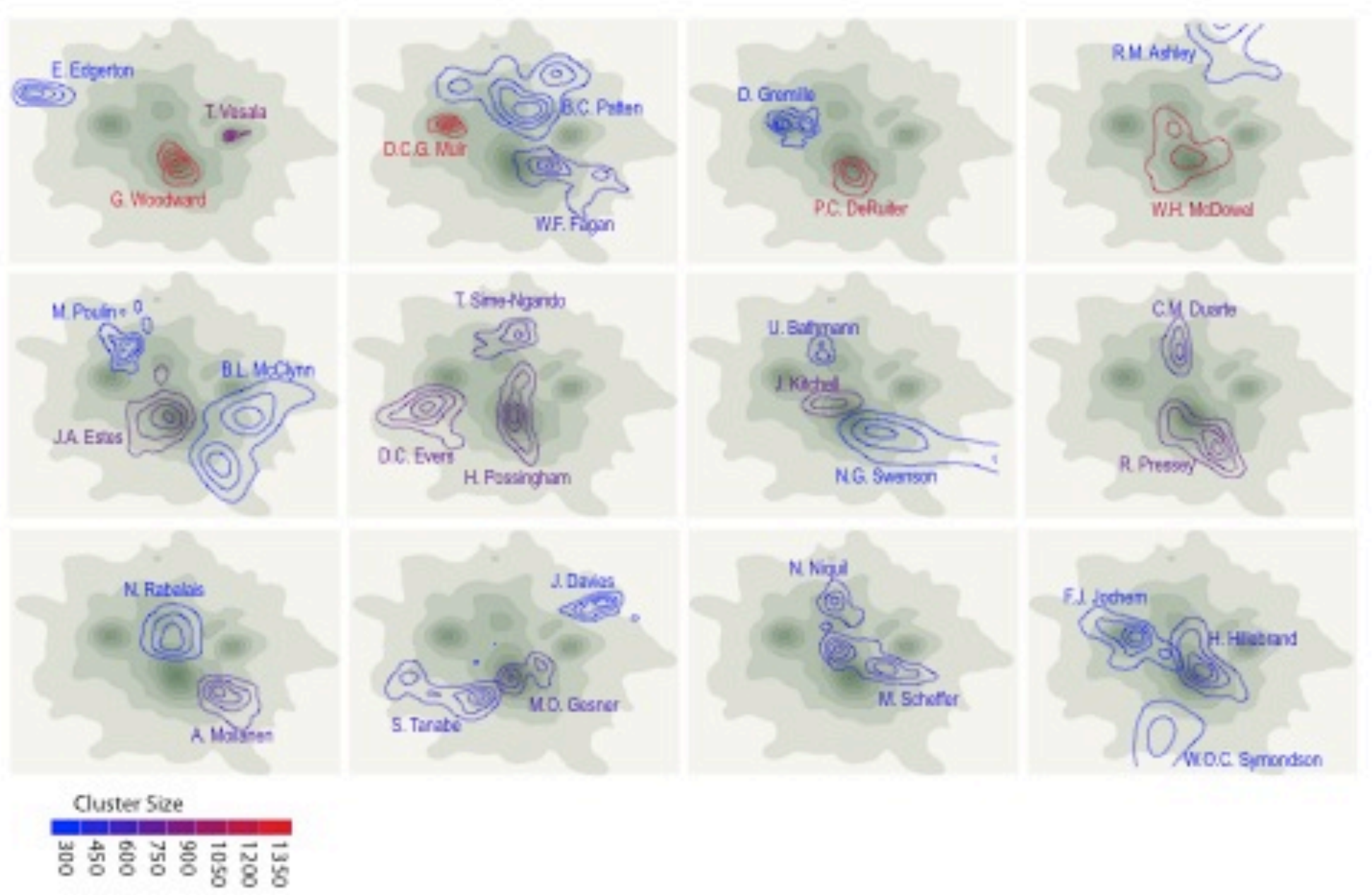

**Figure 8.** Clustering in co-authorship network for Network Ecology publications. Each panel highlights distinct clusters in the co-authorship network and indicates the cluster size, dispersion, and the most central author (betweenness).

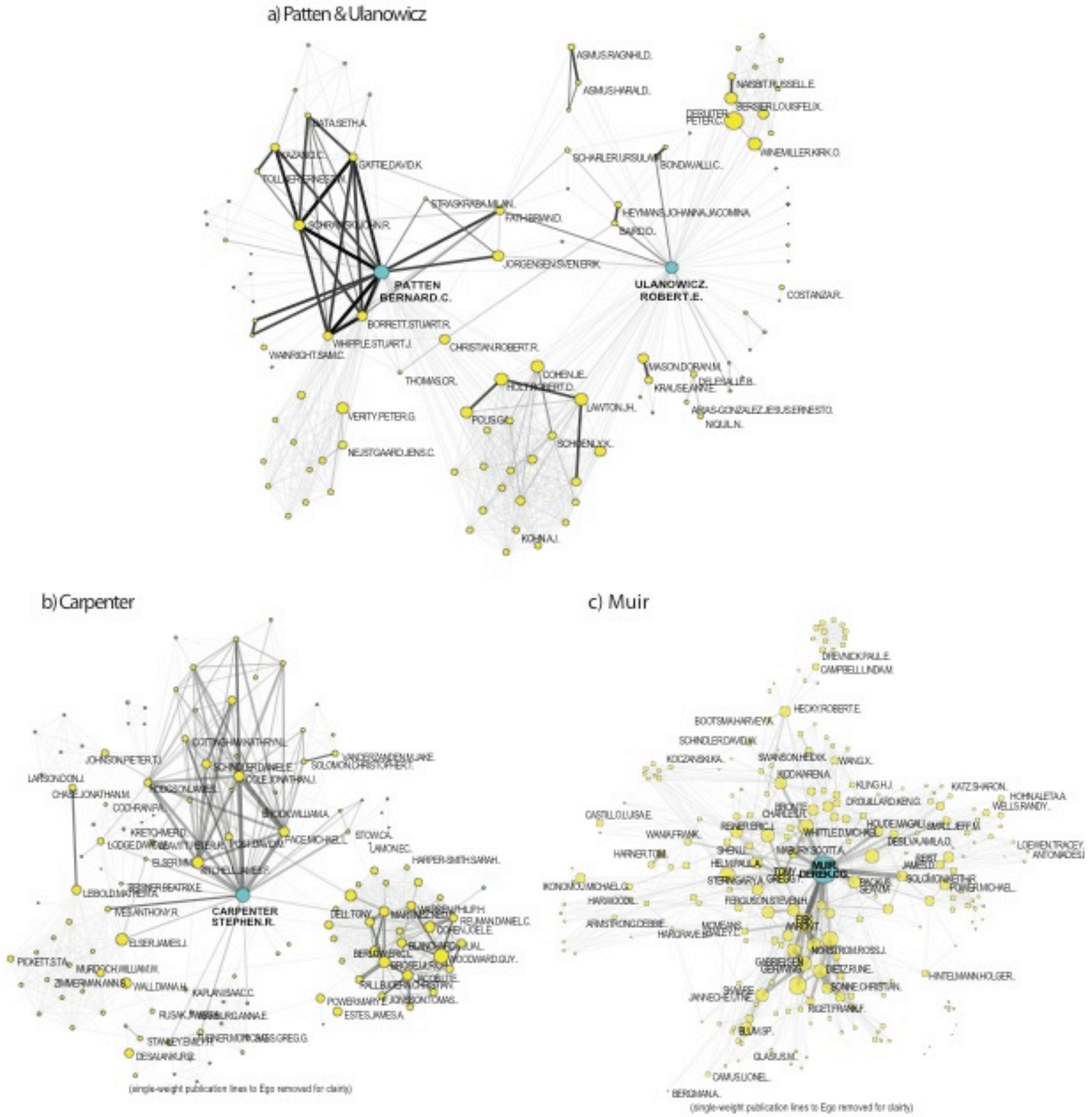

**Figure 9:** One-step ego-networks of (a) Bernard C. Patten and R.E. Ulanowicz, (b) S.R. Carpenter and (c) D.C.G. Muir.

# 10 Online Supplementary Information

**Table S1.** Network ecology survey.

We are currently conducting a study to identify the body of research that uses network models and techniques within ecology. For this purpose, we rely upon your knowledge and expertise. Your responses will be integrated into a review of the network literature. Please help us by answering the *five* questions in this survey. (Participation in this survey is voluntary. We ensure complete anonymity of participants.)

! ! Thank you for your help ! !

1. If you think specifically about the field of ecology within the last 5 years (2007-2012), do you know of any research in that area that has used network models or techniques? By this we mean research that has either *used network analysis / graph theory* as a technique or *employed network concepts* to describe structural characteristics of a phenomena. Please give references for up to 5 publications that you consider most important.

Please indicate some background information on you.

2. Which country are you working in?
   (Drop down list)

3. Where are your currently working in?
   Academia (undergraduate, graduate, postdoc, faculty, non-faculty)
   Private organization / Industry
   Governmental organization
   Other

4. How would you describe the area you are working in?
   (Open ended)

5. Could part of your work be identified as "network ecology"?
   Yes/No